\documentclass[aps,prx,showpacs,twocolumn,nolinenumbers]{revtex4-1}
\UseRawInputEncoding 
\usepackage{upgreek}
\usepackage{hyperref}
\usepackage{amsmath}
\usepackage{amssymb}
\usepackage{amsthm}
\usepackage{mathrsfs} 

\usepackage{graphicx}
\usepackage{xcolor}
\newcommand\mb{\boldsymbol}

\begin{document}

\title{Study on data analysis for Ives-Stilwell-type experiments based on first principles}
\author{Changbiao Wang }
\email{changbiao\_wang@yahoo.com}
\affiliation{ShangGang Group, 10 Dover Road, North Haven, CT 06473, USA}

\begin{abstract}
Ives-Stilwell experiment in 1938 is a historic experiment for confirming Einstein's special relativity, and various modern types have been repeated by use of laser technology.   However in this paper, we reveal and solve a fundamental issue that the data analysis for all those experiments is not consistent with Einstein's definition of the relativistic Doppler effect so that the Doppler effect and its associated time dilation have not actually been confirmed. We argue that there are two first principles for analyzing and confirming Einstein's Doppler effect, stating: (i) Einstein's Doppler effect refers to the same photon (or laser beam) exhibiting different frequencies observed in different inertial frames, and (ii) the quantity (or measurement accuracy) used as a measure to confirm the effect must be able to confirm Einstein's Doppler formula itself.  Unfortunately, Ives-Stilwell data analysis method does not comply with the first principles, failing to find that the experimental data they provided actually does not support Einstein's Doppler effect.  Robertson-Mansouri-Sexl test theory is widely used to test Lorentz invariance, but it does not adhere to the first principles either when employed to measure time dilation.  Based on the first principles, we propose a justified data analysis and correctly confirm the Doppler effect and its associated time dilation in the Ives-Stilwell-type experiment.
\end{abstract}

\maketitle 

\section{Introduction}
\label{s1}
The relativistic Doppler effect is one of the basic results of Einstein's special relativity, and it states that a plane light wave or a single laser beam or a single photon in free space, observed in any inertial frames of reference, propagates at the same speed but it may exhibit different frequencies and aberrations \cite{r1}.  The 1938 Ives-Stilwell experiment \cite{r2} was a historic research work for confirming Einstein's special relativity via the Doppler effect, and various modern types of the experiment \cite{r3,r4,r5,r6,r7} have been repeated by use of laser technology.  The concluding result from an Ives-Stilwell-type experiment is presented in the Letter \cite{r7} by an international group of collaborators after 15 years of efforts \cite{r8}.  However the data analysis method for all those experiments \cite{r2,r3,r4,r5,r6,r7} was not consistent with Einstein's definition of the relativistic Doppler effect so that the Doppler effect and its associated time dilation were not actually confirmed.  For example, the definition of the measurement accuracy of Doppler effect in \cite{r7} is not physical (see Appendix \ref{appa}).  

In this paper, we argue that there are two first principles for analyzing and confirming Einstein's Doppler effect, stating: (i) Einstein's Doppler effect refers to the same photon (or laser beam) exhibiting different frequencies observed in two inertial frames of relative motion, and (ii) the quantity (or measurement accuracy) used as a measure to confirm the effect must be able to confirm Einstein's Doppler formula \emph{itself}.  Unfortunately, Ives and Stilwell performed a data analysis for their experiment test \cite{r2} that did not comply with the first principles so that they even failed to find that the experimental data they provided actually did not support relativistic effect (see Appendix \ref{appb}).  Robertson-Mansouri-Sexl (RMS) test theory is widely used to test Lorentz invariance; however, this RMS theory does not adhere to the first principles either when employed to measure time dilation in Ives-Stilwell-type experiments \cite{r6}(see Appendix \ref{appc}). Here, based on the above first principles with existing experimental data employed \cite{r7}, we propose a justified analysis method and correctly confirm the Doppler effect and its associated time dilation in the Ives-Stilwell-type experiment, thus successfully revealing and solving a long-lasting unrecognized fundamental issue in the experimental confirmation of special relativity.  

\section{Accuracies and data analysis}
\label{s2}
\begin{figure} 
\includegraphics[trim=1.2in 7.0in 1.0in 0.9in, clip=true,scale=0.54]{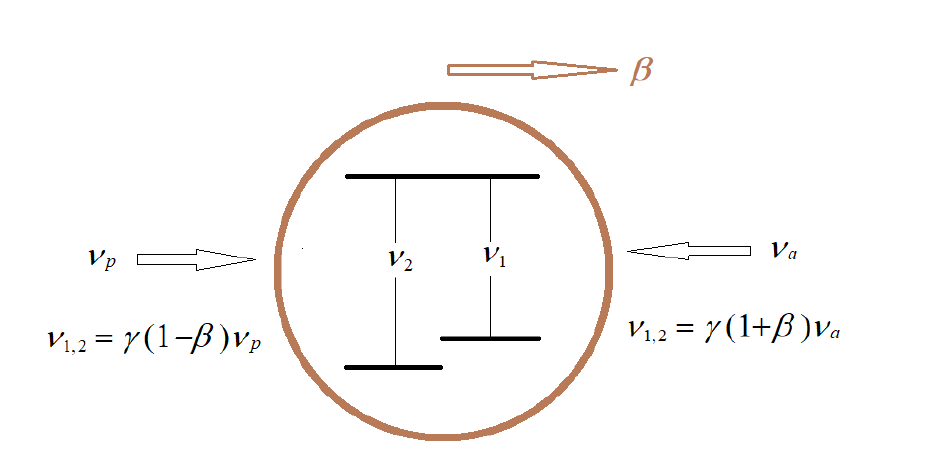}
\caption{Schematic diagram of Ives-Stilwell-type experiment based on optical-optical double resonance (OODR) spectroscopy \cite{r7}.  Two laser beams are aligned antiparallel $(\nu_a)$ and parallel $(\nu_p)$ relative to an ion beam moving at $\beta$.  The ion beam has two transition frequencies $\nu_1$ and $\nu_2$, functioning as Doppler-shifted laser frequency detector.}
\label{fig1}
\end{figure}

In the excellent Letter, Botermann \emph{et al.} \cite{r7} report that the measurement accuracy of relativistic Doppler effect has reached the order of $10^{-9}$ in an Ives-Stilwell-type experiment, where a moving high-energy ion beam is set at resonance with two laser beams aligned parallel and antiparallel relative to the ion beam, and the ions with two very close transition frequencies behave as Doppler-shifted laser frequency detector, as shown in Fig. \ref{fig1}.

However we find that, according to a justified data analysis proposed here, the accuracy obtained by the authors \cite{r7} is not the accuracy for a single laser beam.  The accuracy for a single laser beam is much lower, in the order of $10^{-4}$, different from the accuracy reported in \cite{r7} by five orders of magnitude.  The analysis is given below. 

In their Letter, the definition of measurement accuracy and reported result are given by \cite{r7} 
\begin{align}
\varepsilon&=\sqrt{\frac{\nu_a\nu_p}{\nu_1\nu_2}}-1,  &\textrm{(definition)}
\label{eq1}
\\[5pt]
\varepsilon&=(1.5\pm 2.3)\times 10^{-9},  &\textrm{(reported~ result)}
\label{eq2}
\end{align}
where $\nu_a$  and $\nu_p$  are, respectively, the frequencies of the two laser beams; $\nu_1$  and $\nu_2$  are the transition frequencies of ions; $(\nu_a\nu_p/\nu_1\nu_2)^{1/2}$  is the measured value, with its true value equal to unity.  Mathematically, this definition itself is impeccable, which refers to how a measured value is close to its true value: $accuracy = (x_{\textrm{measured}}-x_{\textrm{true}})/x_{\textrm{true}}$, where $x$ is a physical quantity to be examined.  However physically, what Eq.\,(\ref{eq1}) describes is an expression of the coupling of various accuracies for two individual laser beams, and it cannot represent the accuracy for any single beam of them. Thus Eq.\,(\ref{eq1}) is not consistent with Einstein's definition of the Doppler effect that refers to \emph{a single laser beam} or \emph{a single photon} which, observed in any inertial frames in free space, propagates at the same speed but may exhibit different frequencies and aberrations \cite{r1}. If the measurement accuracy for a single laser beam cannot be obtained, then it is not persuasive and convincing to declare that the Lorentz invariance or special relativity has been verified via the relativistic Doppler effect.

In the Ives-Stilwell-type experiment \cite{r7}, observed in the ion-rest frame there are two measured frequencies (namely transition frequencies):  $\nu_1= 546,455,143.0$ MHz and $\nu_2 = 546,474,960.7$ MHz.  Because the two frequencies are too close plus inevitable systematic errors, it is impossible to identify which transition frequency responds to which laser beam.  Thus we have to evaluate the accuracies respectively for the two measured frequencies for each laser beam.

According to Einstein \cite{r1}, for any measured laser beam frequencies $\nu_a$  and  $\nu_p$, and measured ion's velocity  $\beta c$ (with $c$ the vacuum light speed), we have Doppler-shifted laser frequencies observed in the ion-rest frame, given by
\begin{align}
\nu_{a\textrm{-sh}}&=\gamma(1+\beta)\nu_a,   &\textrm{(antiparallel beam)}
\label{eq3}
\\
\nu_{p\textrm{-sh}}&=\gamma(1-\beta)\nu_p,   &\textrm{(parallel beam)}
\label{eq4}
\end{align}
where $\gamma=(1-\beta^2)^{-1/2}$  is the relativistic factor.    Note that either $\nu_{a\textrm{-sh}}$  or $\nu_{p\textrm{-sh}}$ can be taken as the true value of measured $\nu_1$  or  $\nu_2$.  From Eqs.\,(\ref{eq3}) and (\ref{eq4}) it can be seen that the antiparallel and parallel laser beams are of different types of Doppler effects.

Thus we have two accuracies for measured frequency  $\nu_1$, given by
\begin{align}
\varepsilon_{a1}&=\frac{\nu_1-\nu_{a\textrm{-sh}}}{\nu_{a\textrm{-sh}}},   &\textrm{(antiparallel beam)}
\label{eq5}
\\[5pt]
\varepsilon_{p1}&=\frac{\nu_1-\nu_{p\textrm{-sh}}}{\nu_{p\textrm{-sh}}},    &\textrm{(parallel beam)}
\label{eq6}
\end{align}
and two accuracies for measured frequency  $\nu_2$, given by 
\begin{align}
\varepsilon_{a2}&=\frac{\nu_2-\nu_{a\textrm{-sh}}}{\nu_{a\textrm{-sh}}},   &\textrm{(antiparallel beam)}
\label{eq7}
\\[5pt]
\varepsilon_{p2}&=\frac{\nu_2-\nu_{p\textrm{-sh}}}{\nu_{p\textrm{-sh}}}.    &\textrm{(parallel beam)}
\label{eq8}
\end{align}

From Eqs.\,(\ref{eq3}) and (\ref{eq4}), we have $\nu_a\nu_p=\nu_{a\textrm{-sh}}\nu_{p\textrm{-sh}}$.  Thus Eq.\,(\ref{eq1}) can be written as
\begin{equation}
\varepsilon=\sqrt{\frac{\nu_{a\textrm{-sh}}\nu_{p\textrm{-sh}}}{\nu_1\nu_2}}-1.
\label{eq9}
\end{equation}

From Eq.\,(\ref{eq5}) we have $\nu_1/\nu_{a\textrm{-sh}}=1+\varepsilon_{a1}$  and from Eq.\,(\ref{eq8}) we have $\nu_2/\nu_{p\textrm{-sh}}=1+\varepsilon_{p2}$.  Inserting them into Eq.\,(\ref{eq9}) we have
\begin{equation}
\varepsilon=\frac{1}{\sqrt{ \big(1+\varepsilon_{a1}\big)\big(1+\varepsilon_{p2}\big)} }-1.
\label{eq10}
\end{equation}

From Eq.\,(\ref{eq6}) we have $\nu_1/\nu_{p\textrm{-sh}}=1+\varepsilon_{p1}$  and from Eq.\,(\ref{eq7}) we have $\nu_2/\nu_{a\textrm{-sh}}=1+\varepsilon_{a2}$.  Inserting them into Eq.\,(\ref{eq9}) we have
\begin{equation}
\varepsilon=\frac{1}{\sqrt{ \big(1+\varepsilon_{p1}\big)\big(1+\varepsilon_{a2}\big)} }-1.
\label{eq11}
\end{equation}

According to the Letter \cite{r7}, the measured data are given by $\beta=0.338$, $\nu_1= 546,455,143.0$ MHz, $\nu_2= 546,474,960.7$ MHz, $\nu_a= 384,225,534.98$ MHz ($\lambda_a=780$ nm), and $\nu_p= 777,210,326.98$ MHz ($\lambda_p=386$ nm).   Using these data, from Eqs.\,(\ref{eq3})-(\ref{eq8}) we obtain the accuracies for individual laser beams, given by
\begin{align}
\varepsilon_{a1}&=+3.897,464,375,146\times 10^{-4},
\label{eq12}
\\
\varepsilon_{p1}&=-4.258,480,215,884\times 10^{-4},  
\label{eq13}
\\
\varepsilon_{a2}&=+4.260,264,947,745\times 10^{-4},  
\label{eq14}
\\
\varepsilon_{p2}&=-3.895,975,426,142\times 10^{-4}.  
\label{eq15}
\end{align}

Thus from Eqs.\,(\ref{eq10}) and (\ref{eq11}) we have
\begin{align}
\varepsilon&\approx -\frac{1}{2}(\varepsilon_{a1}+\varepsilon_{p2}+\varepsilon_{a1}\varepsilon_{p2})=1.47\times 10^{-9},
\label{eq16}
\\[5pt]
\varepsilon&\approx-\frac{1}{2}(\varepsilon_{p1}+\varepsilon_{a2}+\varepsilon_{p1}\varepsilon_{a2})=1.47\times 10^{-9}.
\label{eq17}
\end{align}

From Eqs.\,(\ref{eq12})-(\ref{eq15}) we can see that the measurement accuracies of Doppler effect for individual laser beams are in the order of $10^{-4}$, much lower than the accuracy of about  $10^{-9}$ reported in the Letter \cite{r7}.

From Eq.\,(\ref{eq16}) we find that the accuracy  $\varepsilon$, defined in \cite{r7}, is a combination of   $\varepsilon_{a1}$ (the accuracy for antiparallel laser beam for  $\nu_1$) and $\varepsilon_{p2}$  (the accuracy for parallel laser beam for  $\nu_2$).  $\varepsilon_{a1}$  and  $\varepsilon_{p2}$  almost cancel each other to become in the same order of magnitude as  $\varepsilon_{a1}\varepsilon_{p2}$ with opposite signs, namely  $\varepsilon_{a1}+\varepsilon_{p2}=148.8949\times 10^{-9}$ and $\varepsilon_{a1}\varepsilon_{p2}= -151.8443\times 10^{-9}$, so that $\varepsilon=1.47\times10^{-9}$ is obtained.   The same thing happens to Eq.\,(\ref{eq17}). 

In fact, it is easy to understand mathematically that the accuracies for single laser beams are essentially different from the coupling accuracy.  For example, from Eq.\,(\ref{eq10}) we see that even when $\varepsilon = 0$ holds, we can have $\varepsilon_{a1}=-0.5\,(1)$ and $\varepsilon_{p2}= 1\,(-0.5)$ holding, not necessarily  $\varepsilon_{a1}=0$ and $\varepsilon_{p2}=0$. 

One might argue that according to Eqs.\,(\ref{eq1}) and (\ref{eq10})-(\ref{eq17}), the accuracy $\varepsilon=\sqrt{\nu_a\nu_p/(\nu_1\nu_2)}-1$ defined in \cite{r7} is derived from the use of combined quantities to cancel or reduce systematic errors --- a method commonly employed all over experimental science.  However, this is not the case because $\varepsilon=\sqrt{\nu_a\nu_p/(\nu_1\nu_2)}-1$ itself cannot confirm Einstein's Doppler formula even at an ideal situation of $\varepsilon=0$, which has nothing to do with canceling or reducing systematic errors.  

In fact, it is also easy to understand mathematically that $\varepsilon=\sqrt{\nu_a\nu_p/(\nu_1\nu_2)}-1$ cannot confirm Einstein's Doppler formula.  In principle, the validity of Einstein's formulas $\nu_{1}=\gamma(1-\beta)\nu_{p}$ and $\nu_{2}=\gamma(1+\beta)\nu_{a}$ is a sufficient condition for the holding of $\varepsilon=\sqrt{\nu_a\nu_p/(\nu_1\nu_2)}-1=0$, but \emph{not} a necessary condition.  The \emph{sufficient} and \emph{necessary} condition for $\varepsilon=0$ is that $\nu_1\nu_2=\nu_a\nu_p$ holds.  For example, $\nu_{1}=10^{-100}\times\gamma(1-\beta)\nu_{p}$ and $\nu_{2}=10^{100}\times\gamma(1+\beta)\nu_{a}$ also make $\nu_1\nu_2=\nu_a\nu_p$ or $\varepsilon=0$ holding, although the  $\beta$-dependence on $\nu_{1}$ and $\nu_{p}$ or on  $\nu_{2}$ and $\nu_{a}$ is already far from Einstein's Doppler formula in such a case.  From this we can see that $\varepsilon=0$ does not necessarily result from Einstein's Doppler formula.  Therefore, the accuracy $\varepsilon=\sqrt{\nu_a\nu_p/(\nu_1\nu_2)}-1$ defined in \cite{r7} cannot confirm Einstein's Doppler formula itself even at $\varepsilon=0$; see Appendix \ref{appa} for more details.

One might argue that although the quantity  $\varepsilon=\sqrt{\nu_a\nu_p/(\nu_1\nu_2)}-1$ cannot confirm Einstein's Doppler formula, it can be used to test Einstein's time dilation when $\varepsilon=0$ holds.  But this is not true, because $\varepsilon$ itself is completely independent of the time dilation factor  $\gamma$, and one cannot be sure that $\varepsilon=0$ comes from Einstein's relativity; for example, $\nu_{1}=10^{-100}\times\gamma(1-\beta)\nu_{p}$ and $\nu_{2}=10^{100}\times\gamma(1+\beta)\nu_{a}$ also make $\varepsilon=0$ holding, as mentioned above.  Therefore, the quantity $\varepsilon=\sqrt{\nu_a\nu_p/(\nu_1\nu_2)}-1$ defined in \cite{r7} cannot be used to test Einstein's time dilation either. 

However, if Einstein's Doppler formula is confirmed, then its associated time dilation will also be verified.  That is because, as we all know, Einstein's Doppler formula is only different by a time dilation factor from the classic Doppler formula.

\section{Conclusions and remarks}
\label{s3}
In this paper, we have revealed and solved a long-lasting unrecognized fundamental issue on data analysis for Ives-Stilwell-type experiments.  We have shown that $\varepsilon=\sqrt{\nu_a\nu_p/(\nu_1\nu_2)}-1$, given by Eq.\,(\ref{eq1}), is an expression of the coupling of various accuracies for two individual laser beams, and it cannot represent the accuracy for any single beam of them, namely it cannot represent the measurement accuracy of the Doppler effect defined by Einstein.  To confirm this, by use of the experimental data provided in the Letter where the measured ion beam velocity is given by $\beta=0.338$ \cite{r7}, we calculated all four accuracies for single laser beams, showing that the accuracy for a single laser beam is much lower, in the order of $10^{-4}$, different from the accuracy reported in the Letter  by five orders of magnitude; especially, we have proved that the quantity $\varepsilon=\sqrt{\nu_a\nu_p/(\nu_1\nu_2)}-1$ as the definition of the measurement accuracy of Doppler effect in \cite{r7} is unphysical, because Einstein's Doppler formula cannot be confirmed even when $\varepsilon=0$ holds, as shown in Appendix \ref{appa}.

Einstein's definition of the relativistic Doppler effect is widely accepted by the physics community, and it is a scientific consensus, and of course, it is within the scope of the body of knowledge available.

As shown in  Eqs.\,(\ref{eq3}) and (\ref{eq4}), Einstein's Doppler formula includes $\beta$; thus the measurement accuracies for single laser beams proposed in the present paper, given by Eqs.\,(\ref{eq5})-(\ref{eq8}), may sensitively depend on $\beta$, and the $\beta$-value must be known in order to calculate the accuracies.  In contrast, the accuracy $\varepsilon=\sqrt{\nu_a\nu_p/(\nu_1\nu_2)}-1$, defined by Botermann \emph{et al.} \cite{r7}, does not depend on $\beta$, and it is obtained by employing optical-optical double resonance (OODR) spectroscopy, where the antiparallel and parallel laser beams interact with the high-energy ion beam at the same time.  In principle, there is no need to know $\beta$𝛽 to calculate $\varepsilon$ in the OODR technology \cite{r7}.  But $\varepsilon=\sqrt{\nu_a\nu_p/(\nu_1\nu_2)}-1$ is not the accuracy of the Doppler effect for a single laser beam, and it cannot represent the accuracy of the Doppler effect defined by Einstein.  For example, from Eq.\,(\ref{eq11}) we see that even when $\varepsilon = 0$ holds, we can have $\varepsilon_{p1}=1$ and $\varepsilon_{a2}= -0.5$ holding, not necessarily  $\varepsilon_{p1}=0$ and $\varepsilon_{a2}=0$.  Thus one cannot derive $\beta\approx 0.338,377,22$ from $\varepsilon = 0$ $\Rightarrow$ $\varepsilon_{a2}=0$, $\Rightarrow$ $\nu_{a\textrm{-sh}}=\nu_{2}$ due to Eq.\,(\ref{eq7}), $\Rightarrow$ $\nu_{2}=\gamma(1+\beta)\nu_{a}$ due to Eq.\,(\ref{eq3}).  Putting it simply, $\varepsilon = 0$ does not necessarily mean the holding of $\nu_{2}=\gamma(1+\beta)\nu_{a}$ so that $\beta\approx 0.338,377,22$ can be obtained.  Thus the Einstein's Doppler formula  $\nu_{2}=\gamma(1+\beta)\nu_{a}$ is not confirmed even when $\varepsilon = 0$ holds; see Appendix \ref{appa} for more details.

One might criticize that, the analysis in the paper is based mistakenly on an approximate value of the ion beam's velocity ($\beta=0.338$), and the actual velocity of the ions contributing to the [relativistic Doppler] resonance signal can be determined to be $\beta=0.338,377,22$.  If $\beta=0.338,377,22$ is used in the paper, the supposed discrepancy disappears.  

However, if the actual velocity of the ions can be determined by applying the relativistic Doppler formula to the Ives-Stilwell-type experiment, which is designed for confirming the relativistic Doppler formula itself, then the Ives-Stilwell-type experiment must be required to support both the confirmation and application of the Doppler formula at the same time.  This is apparently a \emph{Circular Reasoning} or \emph{Logical Fallacy}.  That is because: (i) to confirm the Doppler formula, it is necessary to assume that the ion velocity $\beta$ has been measured; (ii) conversely, to measure the ion velocity  $\beta$, it is necessary to assume that the Doppler formula has been confirmed.  Thus this criticism is not appropriate.

One might argue that the error reported in the paper simply indicates that $\beta ~[= 0.338,377,22]$ is slightly different from the value $[\beta = 0.338]$ given in the Letter \cite{r7}.  However this is not true.  The error is that the quantity $\varepsilon=\sqrt{\nu_a\nu_p/(\nu_1\nu_2)}-1$ cannot represent the measurement accuracy of Einstein's Doppler effect, because Einstein's Doppler effect refers to the effect of a single laser beam.

It should be noted that, although  $\varepsilon=\sqrt{\nu_a\nu_p/(\nu_1\nu_2)}-1$ is not dependent on $\beta$, the measurement accuracy for a single laser beam is \emph{sensitively} dependent on  $\beta$; for example, $\varepsilon_{a2}=4.260,264,947,745\times 10^{-4}$ for $\beta=0.338$, while $\varepsilon_{a2}=9.432,048,656,75\times 10^{-10}$ for $\beta=0.338,377,22$, with the measurement accuracy changed by five (5) orders of magnitude for a slight $\beta$-change of about $0.1\%$. 

One might argue that although the measurement accuracy of the Doppler effect can only be represented by that from a single laser beam, it is possible to correlate the two measurements from antiparallel and parallel laser beams, which are of different types of Doppler effects, in order to improve accuracy without challenging the physical phenomenon.  However this is not true, because the Doppler phenomenon is judged by checking the accuracy of Doppler effect from a single laser beam, and the correlation of the two measurements may confuse the physics of Doppler effect and change the accuracy, and further challenge the judgment of Doppler phenomenon. 

It should be reiterated that the essential flaw of the Letter \cite{r7} is the definition of the measurement accuracy for Doppler effect inconsistent with Einstein Doppler formula, instead of whether $\beta=0.338$ or $\beta=0.338,377,22$ should be taken in data analysis, although the correct $\beta=0.338$ is taken in the paper. 

Generally speaking, the experiment itself does not prove anything, as it only provides readings of instruments and meters.  It is the researchers of the experiment who use these readings to form a chain of evidence to support their conjectures.  Therefore, for the same readings of instruments and meters, the formed chain of evidence may depend on the way of thinking of the researchers, resulting in different conclusions.  The Ives-Stilwell experiment is probably a typical example, as shown below.

In 1938, Ives and Stilwell famously reported that the results of their experiment ``appear to be a satisfactory confirmation of the Larmor-Lorentz theory'' \cite{r2}, namely their experiment satisfactorily confirmed \emph{Lorentz's ether theory}.  However Jones argued that the predictions obtained from Einstein's special relativity are identical ~with those ~obtained by Ives ~based on Lorentz's ether theory \cite{r9}, namely the Ives-Stilwell experiment satisfactorily confirmed \emph{Einstein's special relativity}. 

From above it can be seen that Jones' and Ives-Stilwell chains of evidence, formed based on the same experimental data, support completely different conclusions. 

The present paper might be another interesting example, with the same experimental data used to form an alternative chain of evidence, leading to a unique and definite conclusion. Although the chain of evidence for single laser beams is quite straightforward, it is the result of the correct way of thinking based on first principles: (i) Einstein's Doppler effect refers to the same photon (laser beam) exhibiting different frequencies observed in two inertial frames of relative motion, and (ii) the quantity (measurement accuracy) used as a measure to confirm the effect must be able to confirm Einstein's Doppler formula itself.  However, the way of thinking set up in the 1938 Ives-Stilwell experiment \cite{r2} was incorrect because it was not consistent with the first principles (see  Appendix \ref{appb}), although it had been mimicked by quite a few generations of physicists \cite{r3,r4,r5,r6,r7}.  Even Einstein seemed unaware of this subtle problem with such a far-reaching impact, which highlights the significance of the first principles proposed in the present paper for analyzing and confirming Einstein's Doppler effect. 

Finally, it should be emphasized that the data analysis method in various Ives-Stilwell-type experiments \cite{r3,r4,r5,r6,r7} is essentially the same as that used by the first Ives-Stilwell experiment in 1938 \cite{r2}, where the measurement accuracies for single light beams were not provided, so in fact, the relativistic Doppler effect was not truly confirmed. Therefore, the present work is the first correct confirmation of the relativistic Doppler effect in the Ives-Stilwell-type experiment based on the data provided by Botermann \emph{et al.} \cite{r7}. 

The principle of relativity requires Maxwell equations to have Lorentz invariance, resulting in invariance of the speed of light in free space \cite{r10} and relativistic Doppler effect \cite{r1}.  Thus the present work is a great advance in the experimental verification of Lorentz invariance via the Doppler effect, just like the verification via the invariance of light speed in the astronomical observations of high-energy gamma rays \cite{r11,r12}. 

\appendix
\section{Einstein's Doppler formula not confirmed at $\varepsilon = 0$}
\label{appa}
In this appendix, by specific calculations based on existing experimental data, we show that the quantity $\varepsilon=\sqrt{\nu_a\nu_p/(\nu_1\nu_2)}-1$ as the definition of the measurement accuracy of Doppler effect in \cite{r7} is physically incorrect, because Einstein's Doppler formula,  $\nu_{2}=\gamma(1+\beta)\nu_{a}$, cannot be confirmed even when  $\varepsilon = 0$ holds.

For $\varepsilon = 0$, from Eq. (\ref{eq1}) and Eq. (\ref{eq11}) we have $\nu_a/\nu_2=\nu_1/\nu_p$ and $(1+\varepsilon_{p1})(1+\varepsilon_{a2})=1$ holding.  In such a case, $\varepsilon_{p1}=(\nu_1-\nu_{p\textrm{-sh}})/\nu_{p\textrm{-sh}}$ and $\varepsilon_{a2}=(\nu_2-\nu_{a\textrm{-sh}})/\nu_{a\textrm{-sh}}$ have the same $\beta$-solution, because the antiparallel and parallel laser beams interact with the same ion beam. 

It should be emphasized that the basic physical requirement for the definition of measurement accuracy of Doppler effect is that Einstein's Doppler formula must be confirmed when the accuracy is equal to zero.  In the single laser beam cases, for example, if $\varepsilon_{p1}=0$, then from Eq. (\ref{eq6}) we have $\nu_1=\nu_{p\textrm{-sh}}$, and from Eq. (\ref{eq4}) we have $\nu_{1}=\gamma(1-\beta)\nu_{p}$, which is the exact Einstein's Doppler formula for the parallel laser beam for the measured frequency $\nu_1$.  Putting it simply, Einstein's Doppler formula $\nu_{1}=\gamma(1-\beta)\nu_{p}$ is strictly confirmed at  $\varepsilon_{p1}=0$.  Similarly, Einstein's Doppler formula $\nu_{2}=\gamma(1+\beta)\nu_{a}$ is confirmed at $\varepsilon_{a2}=0$.  Thus the definition of the measurement accuracy for a single laser beam is physically justified.  However, the definition $\varepsilon=\sqrt{\nu_a\nu_p/(\nu_1\nu_2)}-1$ in \cite{r7} cannot confirm Einstein's Doppler formula at $\varepsilon = 0$, and it is not physical, as shown below.

Inserting Eq. (\ref{eq3}) into Eq. (\ref{eq7}) yields
\begin{equation}
\beta=\frac{\nu^2_2-\nu^2_a(1+\varepsilon_{a2})^2}{\nu^2_2+\nu^2_a(1+\varepsilon_{a2})^2}.
\label{eqa1}
\end{equation}
In the following calculations, the experimental data $\nu_2 = 546,474,960.7$ MHz, $\nu_a = 384,225,534.98$ MHz ($\lambda_a =780$ nm), and $\nu_p=777,210,326.98$ MHz ($\lambda_p=386$ nm) \cite{r7} are taken, while $\nu_1=(\nu_a\nu_p)/\nu_2$ is assumed to make $\varepsilon = 0$ exactly hold.

\begin{enumerate}
\item[(i)] For $\varepsilon = 0$, both $\varepsilon_{p1}=0$ and $\varepsilon_{a2}=0$ can hold.  In such a case, we have $\nu_{2}=\gamma(1+\beta)\nu_{a}$ (Einstein's Doppler formula) holding, and we have
\begin{equation}
\beta=\frac{\nu^2_2-\nu^2_a}{\nu^2_2+\nu^2_a}\approx 0.338,377,220,835.
\label{eqa2}
\end{equation}
\item[(ii)]  For $\varepsilon = 0$,  $\varepsilon_{p1}=1$ and $\varepsilon_{a2}=-0.5$ also can hold, not necessarily  $\varepsilon_{p1}=0$ and $\varepsilon_{a2}=0$.   In such a case we have 
\begin{equation}
\beta=\frac{\nu^2_2-\frac{1}{4}\nu^2_a}{\nu^2_2+\frac{1}{4}\nu^2_a}\approx 0.780,013,866,267,
\label{eqa3}
\end{equation}
greatly deviating from $\beta\approx 0.338,377,220,835$ obtained from Einstein's Doppler formula $\nu_{2}=\gamma(1+\beta)\nu_{a}$ for a single laser beam, as shown in above (i). 
\end{enumerate}
Thus from above (i) and (ii), we can conclude that Einstein's Doppler formula $\nu_{2}=\gamma(1+\beta)\nu_{a}$ is not confirmed at $\varepsilon = 0$, because we cannot be sure that $\nu_{2}=\gamma(1+\beta)\nu_{a}$ will hold when $\varepsilon = 0$ holds.

In fact, from Eq. (\ref{eqa1}) one can see that for given $\nu_2 ~(\ne 0)$ and $\nu_a ~(\ne 0)$, there are an infinite number of $\beta$-values that make $\varepsilon = 0$ hold, where $(-1<\beta<1)$ corresponds to $(+\infty >\varepsilon_{a2}>-1)$.  This can also be understood directly from the definition  $\varepsilon=\sqrt{\nu_a\nu_p/(\nu_1\nu_2)}-1$ itself, where $\varepsilon = 0$ always holds as long as $\nu_a\nu_p/(\nu_1\nu_2)=1$ is true, independent of whether $\nu_i (i=1,2)$ and $\nu_j (j=a,p)$ fulfill Einstein's Doppler formula.

So far we have finished the proof that, the definition $\varepsilon=\sqrt{\nu_a\nu_p/(\nu_1\nu_2)}-1$ in \cite{r7} indeed does not meet the basic physical requirement for the measurement accuracy of Einstein's Doppler effect stated previously, and it is not physical.  

\section{Ives-Stilwell experimental test failing to confirm Einstein's Doppler effect}
\label{appb}

First principles indicates that Einstein's Doppler effect refers to the same photon exhibiting different frequencies observed in different inertial frames, and only the measurement accuracy of Doppler effect from a single light beam can represent the accuracy of Einstein's Doppler effect, because only this accuracy is a measure to confirm Einstein's Doppler formula \emph{itself}.  In this appendix, we will demonstrate that Ives and Stilwell did not provide a data analysis for single light beams in their paper \cite{r2}, and so violated the first principles.  As a result, they were even unware that the experimental data they provided actually did not support Einstein's Doppler effect. 

Ives-Stilwell experimental test was performed with hydrogen canal rays, using the blue-green H$\beta$ line, of wavelength 4861 angstroms \cite{r2}, and the data analysis was conducted based on the \emph{combined} Doppler effect generated by two light beams emitted by moving hydrogen ions in the canal rays.  

In the following, we first examine the data analysis provided by the authors of Ref.\,\cite{r2}, then we propose a unique data analysis for single light beams tailored particularly for the Ives-Stilwell experimental test, and finally, conclusions are given.

\begin{table*} 
\centering
\caption{Experimental data used in Ives-Stilwell analysis, taken from Table I and Table III of Ref.\,\cite{r2}.   $\gamma\beta\lambda_0\approx\beta\lambda_0$ and $(\gamma-1)\lambda_{0}\approx 0.5\beta^2\lambda_{0}$ are, respectively, the true values of the observed quantities  $(\Delta\lambda)_{\beta}$ and $\Delta\lambda$.  The ion's velocity is computed from $\beta=(\beta\lambda_0)/\lambda_0$, with $\lambda_0=4861$$\textrm{\AA}$ the ion's proper wavelength. \vspace{4 pt}} 
\begin{tabular}{|c|c|c|c|c|c|}
\hline
Case/Plate/Line &~$\beta$ ~&~ $\beta\lambda_0$($\textrm{\AA}$) ~&~ $(\Delta\lambda)_{\beta}$($\textrm{\AA}$) ~&~ $0.5\beta^2\lambda_{0}$($\textrm{\AA}$) ~&~ $\Delta\lambda$($\textrm{\AA}$) \\  \hline
1/169/H$_3$ &~ $0.002,184,736$ ~& $10.62$ & $10.35$ & $0.0116$ & $0.011 $ \\ \hline
2/160/H$_2$ &~ $0.002,888,295$ ~& $14.04$ & $14.02$ & $0.0203$ & $0.0185$ \\  \hline
3/163/H$_2$ &~ $0.003,147,501$ ~& $15.30$ & $15.40$ & $0.0238$ & $0.0225$ \\  \hline
4/170/H$_2$ &~ $0.003,361,448$ ~& $16.34$ & $16.49$ & $0.0275$ & $0.027 $ \\  \hline
5/165/H$_3$ &~ $0.002,855,380$ ~& $13.88$ & $14.07$ & $0.0198$ & $0.0205$ \\  \hline
6/172/H$_2$ &~ $0.003,805,801$ ~& $18.50$ & $18.67$ & $0.0352$ & $0.0345$ \\  \hline
7/172/H$_3$ &~ $0.003,096,071$ ~& $15.05$ & $15.14$ & $0.0233$ & $0.0215$ \\  \hline
8/177/H$_2$ &~ $0.004,433,244$ ~& $21.55$ & $21.37$ & $0.0478$ & $0.047$  \\  \hline
\end{tabular}
\label{tab1}
\end{table*}

\textbf{Ives-Stilwell data analysis.}  The second-order combined Doppler effect is described by
\begin{equation}
\frac{1}{2}(\lambda_{b}+\lambda_{f})-\lambda_{0}=(\gamma-1)\lambda_{0},
\label{eqb1}
\end{equation}
where $\gamma$ is the ion's Lorentz factor, and $\lambda_0$ is its proper wavelength; $\lambda_{b}$ and $\lambda_{f}$ are, respectively, the wavelengths of light emitted by the ions in backward and forward directions, observed in the laboratory frame. 
If both the dependence of $\lambda_{b}$ and the dependence of $\lambda_{f}$ on $\lambda_{0}$ respectively fulfill Einstein's Doppler formula, then Eq.\,(\ref{eqb1}) holds exactly.
 
In Ives-Stilwell data analysis \cite{r2}, the left side of Eq.\,(\ref{eqb1}) given by
\begin{equation}
 \Delta\lambda\equiv \frac{1}{2}(\lambda_{b}+\lambda_{f})-\lambda_{0}
\label{eqb2}
\end{equation}
is a quantity to be measured (or observed), which means that all $\lambda_{b}$, $\lambda_{f}$, and $\lambda_{0}$ are measured and they may not exactly satisfy Einstein's Doppler formula, thus causing a deviation from Eq.\,(\ref{eqb1}).  To examine the deviation, the right side of Eq.\,(\ref{eqb1}) given by
\begin{equation}
 (\gamma-1)\lambda_{0}\approx\frac{1}{2}\beta^2\lambda_{0} \quad (\sim\beta^2, ~\textrm{second-order})
\label{eqb3}
\end{equation}
is taken as the true value of $\Delta\lambda$, where $\beta$ is the ion's velocity normalized to vacuum light speed.  Thus the quantity or measurement accuracy
\begin{equation}
 \varepsilon=\frac{\Delta\lambda}{(\gamma-1)\lambda_{0}}-1 
\label{eqb4}
\end{equation}
is a measure to confirm the second-order combined Doppler effect described by Eq.\,(\ref{eqb1}), because Eq.\,(\ref{eqb1}) exactly holds at $\varepsilon=0$, while $\varepsilon\neq 0$ means a deviation from the second-order \emph{combined} Doppler effect (instead of a \emph{single-beam} Doppler effect).

\begin{table} 
\centering
\caption{Accuracies of first-order and second-order \emph{combined} Doppler effect, computed from the data in Table \ref{tab1}.  The accuracies of first-order effect have a range of $-2.6\%<\varepsilon_{\beta}<1.4\%$ while the accuracies of second-order effect have a range of $-8.9\%<\varepsilon<3.6\%$.  Note that if a true value is larger than the value of its observed quantity, then the accuracy is negative; otherwise it is positive.\vspace{4 pt} } 
\begin{tabular}{|c|c|c|}
\hline
Case/Plate/Line &~$\varepsilon_{\beta}$ ~&~ ~$\varepsilon$  \\  \hline
1/169/H$_3$ &~ $-0.0254$ ~&~ $-0.0517$~  \\ \hline
2/160/H$_2$ &~ $-0.0014$ ~&~ $-0.0887$~  \\ \hline
3/163/H$_2$ &~ $+0.0065$ ~&~ $-0.0546$~  \\  \hline
4/170/H$_2$ &~ $+0.0092$ ~&~ $-0.0182$~  \\  \hline
5/165/H$_3$ &~ $+0.0137$ ~&~ $+0.0354$~  \\  \hline
6/172/H$_2$ &~ $+0.0092$ ~&~ $-0.0199$~  \\  \hline
7/172/H$_3$ &~ $+0.0060$ ~&~ $-0.0773$~  \\  \hline
8/177/H$_2$ &~ $-0.0084$ ~&~ $-0.0167$~  \\  \hline
\end{tabular}
\label{tab2}
\end{table}

The first-order combined Doppler effect is described by 
\begin{equation} 
\frac{1}{2\cos{\theta}}(\lambda_b-\lambda_f)=\gamma\beta\lambda_0, \vspace{6 pt}
\label{eqb5}
\end{equation}							
where $\theta=7^{\circ}$ is the observation angle.   If the dependence of $\lambda_b$ and $\lambda_f$ on $\lambda_0$ respectively fulfill Einstein's Doppler formula, then Eq.\,(\ref{eqb5}) holds exactly.

In their analysis \cite{r2}, the left side of Eq.\,(\ref{eqb5}) given by
\begin{equation}
(\Delta\lambda)_{\beta}=\frac{1}{2\cos{\theta}}(\lambda_b-\lambda_f)
\label{eqb6}
\end{equation}	
is a quantity to be measured (or observed), which means that measured $\lambda_{b}$, $\lambda_{f}$, and $\lambda_{0}$  may result in a deviation from Eq.\,(\ref{eqb5}).  To examine the deviation, the right side of Eq.\,(\ref{eqb5}) given by
\begin{align}
\gamma\beta\lambda_0&\approx\Big(1+\frac{1}{2}\beta^2\Big)\beta\lambda_0 \nonumber  \\
&\approx \beta\lambda_0 \quad (\sim\beta, ~\textrm{first-order})
\label{eqb7}
\end{align}	
is taken as the true value of $(\Delta\lambda)_{\beta}$.  Thus the quantity or measurement accuracy
\begin{equation}
\varepsilon_{\beta}=\frac{(\Delta\lambda)_{\beta}}{\gamma\beta\lambda_0}-1
\label{eqb8}
\end{equation}	
is a measure to confirm the first-order combined Doppler effect described by Eq.\,(\ref{eqb5}), because Eq.\,(\ref{eqb5}) exactly holds at $\varepsilon_{\beta}=0$, while $\varepsilon_{\beta}\neq 0$ means a deviation from the first-order \emph{combined} Doppler effect (instead of a \emph{single-beam} Doppler effect).

It should be emphasized that physically, $\varepsilon_{\beta}=0$ or Eq.\,(\ref{eqb5}) cannot be used to get the ion's velocity $\beta$, because Eq.\,(\ref{eqb5}), just like Eq.\,(\ref{eqb1}), is \emph{to be confirmed}; otherwise, circular reasoning would result.  Nevertheless, Ives-Stilwell analysis \cite{r2} seemed to have provided this treatment in column 5 of Table III for $0.5\beta^2\lambda_0$ as a comparison, where $\beta$ was obtained from $\varepsilon_{\beta}=0\Rightarrow (\Delta\lambda)_{\beta}=\gamma\beta\lambda_0\approx\beta\lambda_0\Rightarrow \beta\approx(\Delta\lambda)_{\beta}/\lambda_0$.  But such a treatment is not justified if the purpose is to avoid measuring $\beta$, because it is really a logic fallacy. 
 
The authors of Ref.\,\cite{r2} did not provide measured wavelengths explicitly but they can be obtained from Eqs.\,(\ref{eqb2}) and (\ref{eqb6}), given by
\begin{align}
&\lambda_b=\lambda_0+\Delta\lambda+(\Delta\lambda)_{\beta}\cos{\theta} \quad \textrm{(backward light)}, 
\label{eqb9}
\\
&\lambda_f=\lambda_0+\Delta\lambda-(\Delta\lambda)_{\beta}\cos{\theta} \quad \textrm{(forward light)}. 
\label{eqb10}
\end{align}
The above measured wavelengths will be used to conduct the data analysis for single light beams.

The experimental data that Ref.\,\cite{r2} provided are shown in Table \ref{tab1} and the accuracies derived from the data are shown in Table \ref{tab2}.  (Note that the ion's velocity $\beta$ is derived from $\beta=(\beta\lambda_0)/\lambda_0$, with $(\beta\lambda_0)$ taken from Table I of Ref.\,\cite{r2}, but $(\beta\lambda_0)$ there is only rounded to the second decimal place, which may cause numerical errors for $0.5\beta^2\lambda_0$ given in column 4 in Table III of Ref.\,\cite{r2}.  For example, $\beta=(\beta\lambda_0)/\lambda_0=15.30/4861\approx 0.003147501$ for case 3, and we have  $0.5\beta^2\lambda_0\approx 0.0241$, compared to  $0.5\beta^2\lambda_0=0.0238$ given in column 4 in Table III of Ref.\,\cite{r2}.  But this error would have no essential effect.)

From Tables \ref{tab1} and \ref{tab2}, we can see that Ives-Stilwell analysis \cite{r2} aimed to check how the observed quantities $\Delta\lambda$ and $(\Delta\lambda)_{\beta}$ are close to their true values $(\gamma-1)\lambda_0\approx 0.5\beta^2\lambda_0$ and $\gamma\beta\lambda_0\approx\beta\lambda_0$, respectively.  According to $\Delta\lambda=0.5\beta^2\lambda_0$ and $\Delta\lambda=[(\Delta\lambda)_{\beta}]^2/(2\lambda_0)$, as shown in Fig.\,13 and Fig.\,14 of Ref.\,\cite{r2}, the authors concluded that the results exhibiting in the figures ``appear to be a satisfactory confirmation of the Larmor-Lorentz theory [Einstein's special theory]''. However $\Delta\lambda=0.5\beta^2\lambda_0$ and $\Delta\lambda=[(\Delta\lambda)_{\beta}]^2/(2\lambda_0)$ were obtained by approximate treatment of the true values: $(\gamma-1)\lambda_0\approx 0.5\beta^2\lambda_0$ and $\gamma\beta\lambda_0\approx\beta\lambda_0$, as shown in Eqs.\,(\ref{eqb3}) and (\ref{eqb7}), and whether they can be used as a criterion is questionable, which can be clearly seen from the data analysis for single light beams shown below.  

\textbf{Data analysis for single light beams.}  According to Einstein \cite{r1}, the Doppler shifted wavelengths for the two light beams are, respectively, given by 
\begin{align}
&\lambda_{b-\textrm{sh}}=\gamma(1+\beta\cos{\theta})\lambda_{0} &&\textrm{for backward light},
\label{eqb11}
\\[5pt]
&\lambda_{f-\textrm{sh}}=\gamma(1-\beta\cos{\theta})\lambda_{0} &&\textrm{for forward light},
\label{eqb12}
\end{align}
where $\theta = 7^{\circ}$ is the observation angle; $\lambda_{0}$ is the photon's wavelength observed in the ion's rest frame, as mentioned before, while $\lambda_{b-\textrm{sh}}$ or $\lambda_{f-\textrm{sh}}$ is the wavelength of the \emph{same} photon observed in the laboratory frame, which constitutes one of the first principles: Einstein's Doppler effect refers to the same photon exhibiting different frequencies observed in different inertial frames.

From above, the measurement accuracies of Einstein's Doppler effect for single light beams are given by
\begin{align}
&\varepsilon_{b}=\frac{\lambda_{b}-\lambda_{b-\textrm{sh}}}{\lambda_{b-\textrm{sh}}}  &&\textrm{for backward light},
\label{eqb13}
\\[5pt]
&\varepsilon_{f}=\frac{\lambda_{f}-\lambda_{f-\textrm{sh}}}{\lambda_{f-\textrm{sh}}}  &&\textrm{for forward light},
\label{eqb14}
\end{align}
where $\lambda_{b-\textrm{sh}}$ is the theoretically expected value, taken as the true value of the measured wavelength $\lambda_{b}$, while $\lambda_{f-\textrm{sh}}$ is the true value of the measured $\lambda_{f}$.

The accuracies  $\varepsilon_{b}$ and $\varepsilon_{f}$ have clear physical meaning.  For example, if $\varepsilon_{b}=0$ holds, then we have $\lambda_{b}=\lambda_{b-\textrm{sh}}$ and the Doppler effect of a single photon described by Einstein's formula Eq.\,(\ref{eqb11}) is exactly confirmed.  In general, $\varepsilon_{b} ~(\neq 0)$ denotes a deviation from the formula for this photon.  Thus the quantity $\varepsilon_{b}$ is a measure to confirm Einstein's formula, and only the Doppler effect from a single photon can represent Einstein's Doppler effect, which constitutes the other one of the first principles: the quantity (or accuracy) used as a measure to confirm Einstein's Doppler effect must be able to confirm Einstein's Doppler formula itself.  

From Eqs.(\ref{eqb13}) and  (\ref{eqb14}) we have  
\begin{align}
&\lambda_{b}=(1+\varepsilon_{b})\gamma(1+\beta\cos{\theta})\lambda_{0} &&\textrm{(backward)},
\label{eqb15}
\\[5pt]
&\lambda_{f}=(1+\varepsilon_{f})\gamma(1-\beta\cos{\theta})\lambda_{0} &&\textrm{(forward)}.
\label{eqb16}
\end{align}

According to above Eqs.\,(\ref{eqb15}) and (\ref{eqb16}), we can draw a useful criterion, stating that a \emph{necessary} condition for observing the relativistic Doppler effect from the backward or forward beam is that
\begin{equation}
\varepsilon_b>\varepsilon_c \quad\textrm{or}\quad \varepsilon_f>\varepsilon_c \quad\textrm{with}\quad \varepsilon_c=\frac{1}{\gamma}-1
\label{eqb17}
\end{equation}
must hold, where $\varepsilon_c$ is the critical accuracy. That is because, for example, if $(1+\varepsilon_b)\gamma=1$ holds, then we have $\lambda_{b}=(1+\beta\cos{\theta})\lambda_{0}$ which is the classical Doppler wavelength.  But the relativistic wavelength must be larger than its corresponding expected classic wavelength.  Thus $(1+\varepsilon_b)\gamma>1$ must hold for Einstein's Doppler effect, leading to $\varepsilon_b>\varepsilon_c$.  The same argument applies to $\varepsilon_f$.  Thus the holding of $\varepsilon_b>\varepsilon_c$ ($\varepsilon_f>\varepsilon_c$) is a necessary condition to support Einstein's Doppler effect for the backward (forward) light beam.  This simple criterion will be used to examine Ives-Stilwell experimental test \cite{r2}, as shown below.

Inserting Eq.\,(\ref{eqb15}) into Eq.\,(\ref{eqb9}) and Eq.\,(\ref{eqb16}) into Eq.\,(\ref{eqb10}), we obtain 
\begin{align}
\varepsilon_b&=\frac{D_b}{ \gamma(1+\beta\cos{\theta})\lambda_0 }+\varepsilon_c, 
\label{eqb18}
\\ 
\varepsilon_f&=\frac{D_f}{ \gamma(1-\beta\cos{\theta})\lambda_0 }+\varepsilon_c, 
\label{eqb19}
\end{align}
where $D_b$ and $D_f$ are the discriminants, given by

\begin{align}
D_b=\Delta\lambda+\big[(\Delta\lambda)_{\beta}-\beta\lambda_0\big]\cos{\theta}, 
\label{eqb20}
\\
D_f=\Delta\lambda-\big[(\Delta\lambda)_{\beta}-\beta\lambda_0\big]\cos{\theta}.
\label{eqb21}
\end{align} 

Note that (i) in above Eqs.\,(\ref{eqb18}) and (\ref{eqb19}), $D_b$, $D_f$ and $\lambda_0$ can be treated as being dimensionless for simplicity and clarity, because they have the same dimension and show up in the form of $D_b/\lambda_0$ and $D_f/\lambda_0$, so the dimension can be considered to be cancelled each other; (ii) no approximate treatment is invoked in obtaining $D_b$ and $D_f$.

If $D_b$  ($D_f$)$<0$ holds, then we have $\varepsilon_b$ ($\varepsilon_f$)$<\varepsilon_c$, because both $\gamma(1+\beta\cos{\theta})\lambda_0>0$ and $\gamma(1-\beta\cos{\theta})\lambda_0>0$ hold.  Physically,  $\varepsilon_b$ ($\varepsilon_f$)$<\varepsilon_c$ means that the measured wavelength $\lambda_b$ ($\lambda_f$) is less than its classic Doppler wavelength, so there is no expected relativistic effect taking place.  

Inserting the data in Table \ref{tab1} into Eqs.\,(\ref{eqb20}) and (\ref{eqb21}), we obtain all the values of the discriminants $D_b$ and $D_f$, as shown in Table \ref{tab3}. 

From Table \ref{tab3}, we find that for each of all 8 cases in Ives-Stilwell experimental test, we have either $D_b<0$ or $D_f<0$ holding, so at least one of the two light beams does not have the relativistic effect predicted by Einstein's special theory, a result contrary to that claimed by the authors of Ref.\,\cite{r2}, which means that the relation expressions $\Delta\lambda=0.5\beta^2\lambda_0$ and $\Delta\lambda=[(\Delta\lambda)_{\beta}]^2/(2\lambda_0)$  used as a criterion in \cite{r2} is not correct.  Thus we can conclude that Ives-Stilwell data analysis failed to confirm Einstein's Doppler effect. 

\begin{table} 
\centering
\caption{Numerical results of discriminants $D_b$ and $D_f$, computed from the data in Table \ref{tab1}.  If $D_b$ ($D_f$)$<0$, then $\varepsilon_b$ ($\varepsilon_f$) $<\varepsilon_c$, and the measured wavelength $\lambda_b$ ($\lambda_f$) is less than its classic Doppler wavelength, with no expected relativistic effect occurring.\vspace{4 pt} } 
\begin{tabular}{|c|c|c|}
\hline
Case/Plate/Line &~$D_b$ ~&~ ~$D_f$  \\  \hline
1/169/H$_3$ &~ $-0.2570$ ~&~ $+0.2790$~  \\  \hline
2/160/H$_2$ &~ $-0.0014$ ~&~ $+0.0384$~  \\  \hline
3/163/H$_2$ &~ $+0.1218$ ~&~ $-0.0768$~  \\  \hline
4/170/H$_2$ &~ $+0.1759$ ~&~ $-0.1219$~  \\  \hline
5/165/H$_3$ &~ $+0.2091$ ~&~ $-0.1681$~  \\  \hline
6/172/H$_2$ &~ $+0.2032$ ~&~ $-0.1342$~  \\  \hline
7/172/H$_3$ &~ $+0.1108$ ~&~ $-0.0678$~  \\  \hline
8/177/H$_2$ &~ $-0.1317$ ~&~ $+0.2257$~  \\  \hline
\end{tabular}
\label{tab3}
\end{table}

Einstein's relativistic Doppler effect for a plane light wave is described by \cite{r1}
\begin{equation}
\lambda_{\textrm{lab}}=\gamma(1-\hat{\mathbf{n}}\cdot\mb{\beta})\lambda_0,
\label{eqb22}
\end{equation}
where  $\lambda_0$  is the wavelength observed in the moving frame, such as in the ion-rest frame in the Ives-Stilwell case, $\lambda_{\textrm{lab}}$  is the wavelength observed in the laboratory frame, $\mb{\beta}$ is the velocity at which the frame moves relatively to the laboratory frame, and $\hat{\mathbf{n}}$ is the unit wave vector.  From this we can see that the relativistic Doppler effect exists observed in any directions in the laboratory frame as long as $\mb{\beta}\neq 0$ or $\gamma=(1-\boldsymbol{\beta}^2)^{-1/2}>1$ holds.  However in the Ives-Stilwell experiment test, out of all 8 cases, each has a light beam that does not have the relativistic effect.  For example, for case 8 with $D_b = -0.1317 <0 $ leading to $\varepsilon_b = -3.7\times 10^{-5} < \varepsilon_c = -9.8\times 10^{-6}$, the backward beam has a measured wavelength of $4882.26$$\textrm{\AA}$ less than its classic Doppler wavelength of $4882.39$$\textrm{\AA}$, without any relativistic effect. (See Ref. \cite{r15} for all detailed numerical analysis of the Ives-Stilwell experimental test.)

From above, we can conclude that the experimental test itself failed to confirm the existence of Einstein's Doppler effect.

\textbf{Conclusions.}  We have demonstrated that Ives and Stilwell did not provide a data analysis for individual light beams in their paper \cite{r2}.  Their analysis was based on the combined Doppler effect from two light beams generated by hydrogen ions in canal rays, and took the relation expressions $\Delta\lambda=0.5\beta^2\lambda_0$ and $\Delta\lambda=[(\Delta\lambda)_{\beta}]^2/(2\lambda_0)$ as a criterion.  However this was incorrect, as shown in Table \ref{tab3}. Thus Ives-Stilwell data analysis failed to confirm Einstein's Doppler effect.  

According to first principles, we have proposed a justified unique data analysis to confirm Einstein's Doppler effect for Ives-Stilwell experimental test \cite{r2}.  We have strictly derived a simple criterion and two discriminants, given by Eq.\,(\ref{eqb17}), and Eqs.\,(\ref{eqb20}) and (\ref{eqb21}), which are tailored to analyze the data provided by the experimental test \cite{r2}, finding that out of all 8 cases, there is no one to support the relativistic Doppler effect, as shown in Table \ref{tab3}.  

Therefore, we conclude that Ives-Stilwell data analysis and the experimental test itself both failed to confirm Einstein's Doppler effect. 

\section{Time dilation failed to be confirmed in the optical-atomic-clock experiment}
\label{appc}
Reinhardt and coworkers \cite{r6} reported that time dilation is confirmed to unprecedented precision with a Mansouri-Sexl parameter $|\alpha|\le8.4 \times 10^{-8}$ in an Ives-Stilwell-type experiment based on fast optical atomic clocks.  However, we would like to point out that the data analysis method for this experiment is flawed, and thus the conclusion obtained in the paper \cite{r6} is under question, which is shown below.

In this paper \cite{r6}, a test theory by Robertson, Mansouri and Sexl (RMS) is adopted to analyze their experiment, where RMS transformation is a generalized Lorentz transformation, and the RMS transformation of the Ives-Stilwell experiment is modified as 
\begin{equation}
\frac{\nu_{a}\nu_{p}}{\nu_{0}^2}=1+2\alpha\beta^2,
\label{eqc1}
\end{equation} 
where $\alpha$ is the Mansouri-Sexl parameter;  $\nu_a$  and $\nu_p$  are, respectively, the frequencies of two laser beams; $\nu_0$ is the transition frequency of ions; $\beta$ is the ion's moving velocity.  

According to this theory, the time dilation will be confirmed when 
\begin{equation}
\alpha=0
\label{eqc2}
\end{equation}
holds. Unfortunately, the RMS theory is not applicable in such a case, unless the authors, Reinhardt and coworkers \cite{r6}, assume that $\nu_{a}\nu_{p}=\nu_{0}^2$ only comes from Einstein's Doppler formula
\begin{equation}
\nu_{p,a}=\gamma(1\pm\beta)\nu_{0}, 
\label{eqc3}
\end{equation} 
as done in \cite{r13}, or that the relation $\nu_{a}\nu_{p}=\nu_{0}^2$ is valid only in special relativity, as claimed by L\"{a}mmerzahl \cite{r14}.  In fact, however, even if the special relativity does not hold or Lorentz invariance is violated,
\begin{equation}
\nu_{p,a}=10^{\pm 100}\gamma(1\pm\beta)\nu_{0}
\label{eqc4}
\end{equation}
for example, we also can get the $\beta$-independent relation $\nu_{a}\nu_{p}=\nu_{0}^2$.  Thus even if $\nu_{a}\nu_{p}=\nu_{0}^2$ or $\alpha=0$ exactly holds, Einstein's Doppler formula or time dilation cannot be confirmed. 

The failure of RMS test theory in this paper \cite{r6} lies in the fact that the Mansouri-Sexl parameter $\alpha$ is designed to measure time dilation via the \emph{combined} Doppler effect from two different laser beams, instead of the effect from a single laser beam, which is in violation of first principles: (i) Einstein's Doppler effect refers to the same photon exhibiting different frequencies observed in different inertial frames, and (ii) the quantity used as a measure to confirm the Doppler effect must be able to confirm Einstein's Doppler formula \emph{itself}.

\begin{description}
\item [Acknowledgments]  The author acknowledges the paper by Ives and Stilwell \cite{r2} and the Letter by Botermann \emph{et al.} \cite{r7}, without which the present work would not have been possible. Note: Like citing published theoretical formulas (Einstein's Doppler formula, for example), it is allowable and legitimate to cite the experimental data in \cite{r2} and \cite{r7} to support the theory developed in the present paper, and it is not in violation of any ethics guidelines.
\item [Disclosures] The author declares no conflicts of interest. 
\item [Data availability] Data underlying the results presented in this paper are available within the paper and Supplement 1 \cite{r15}.    
\end{description}


\newpage
\begin{figure} 
\includegraphics[trim=1.0in 1.0in 1.0in 1.0in, clip=true,scale=1.0]{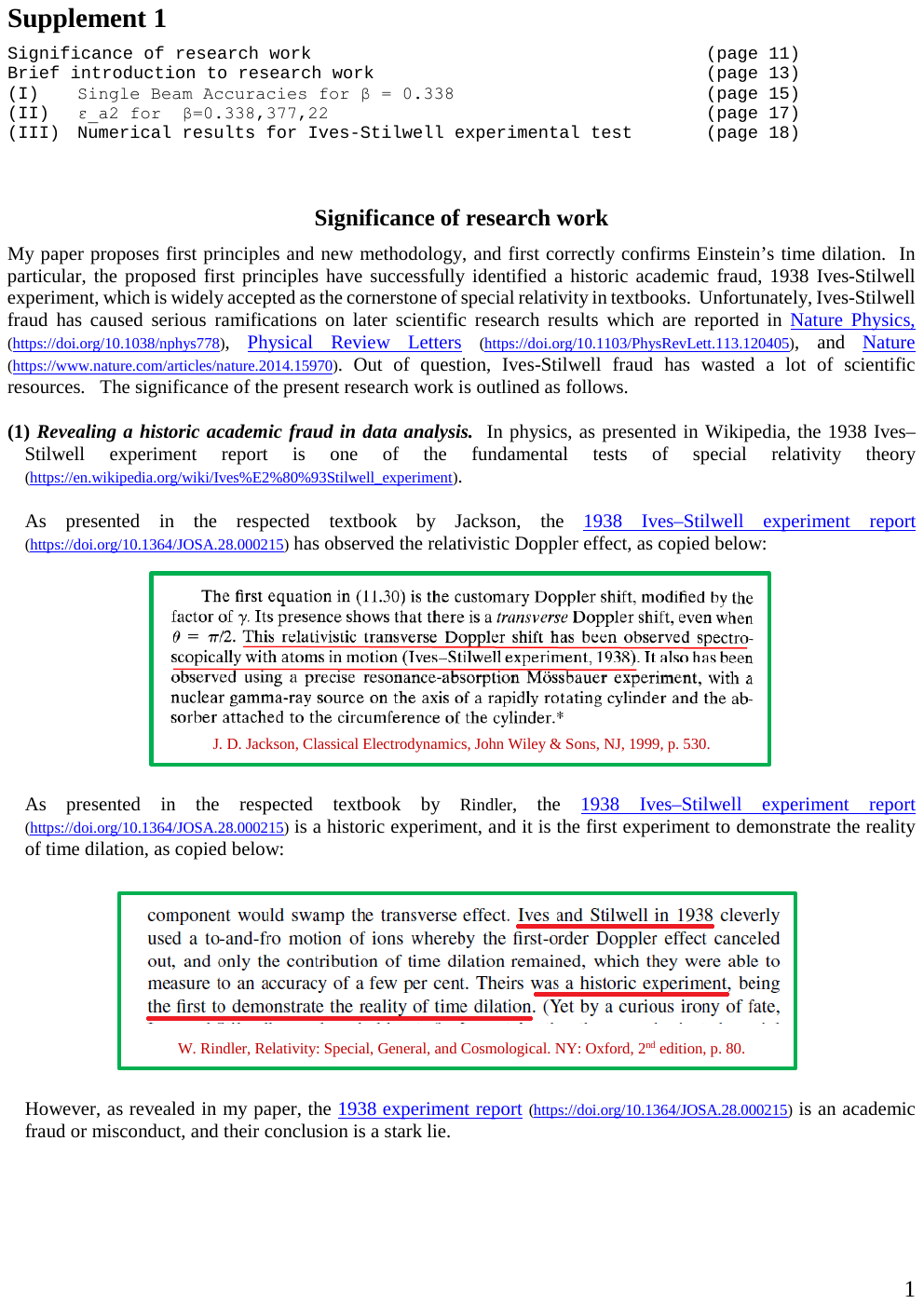}
\label{figM1}
\end{figure} 

\begin{figure} 
\includegraphics[trim=1.0in 1.0in 1.0in 1.0in, clip=true,scale=1.0]{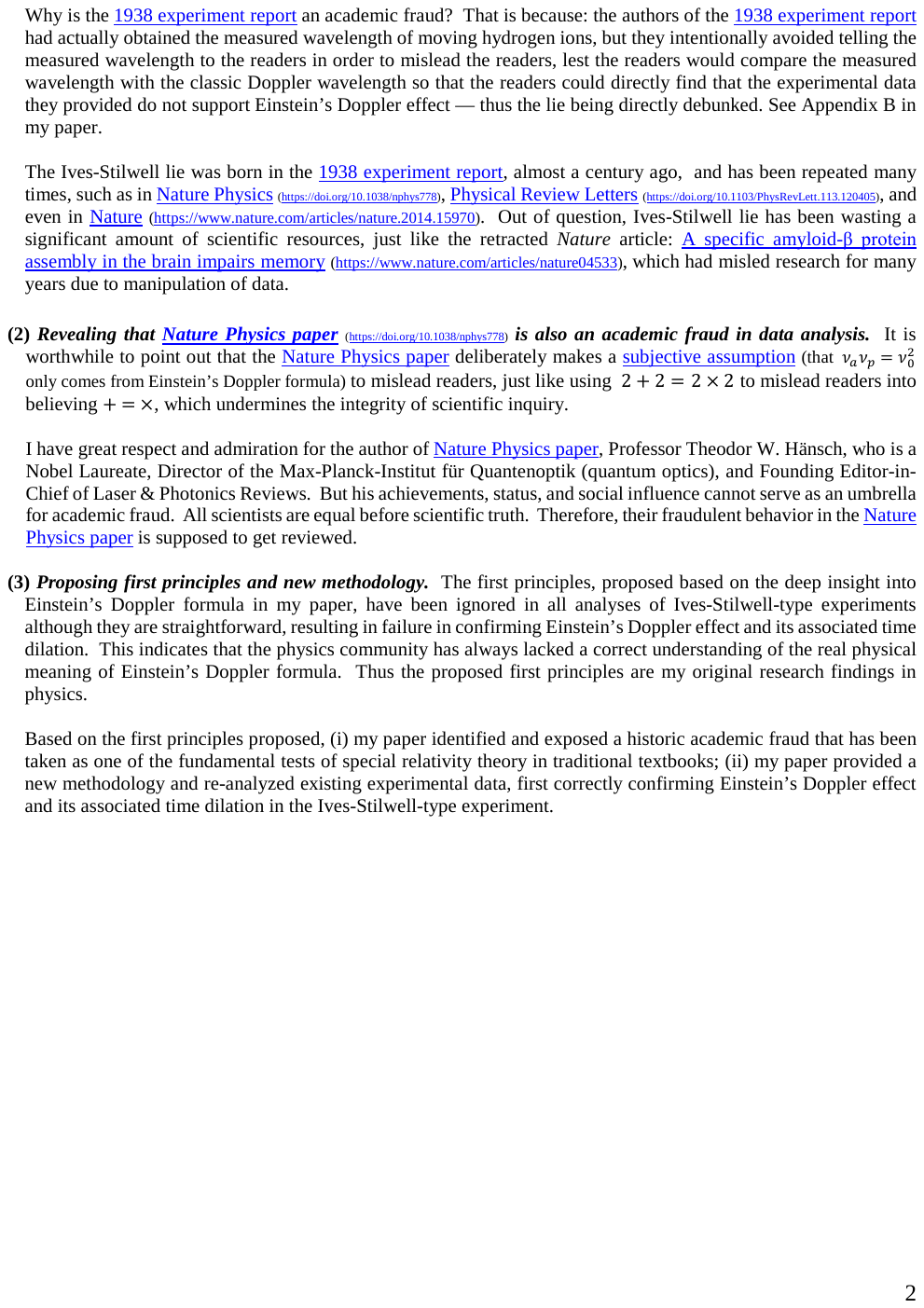}
\label{figM2}
\end{figure} 

\begin{figure} 
\includegraphics[trim=1.0in 1.0in 1.0in 1.0in, clip=true,scale=1.0]{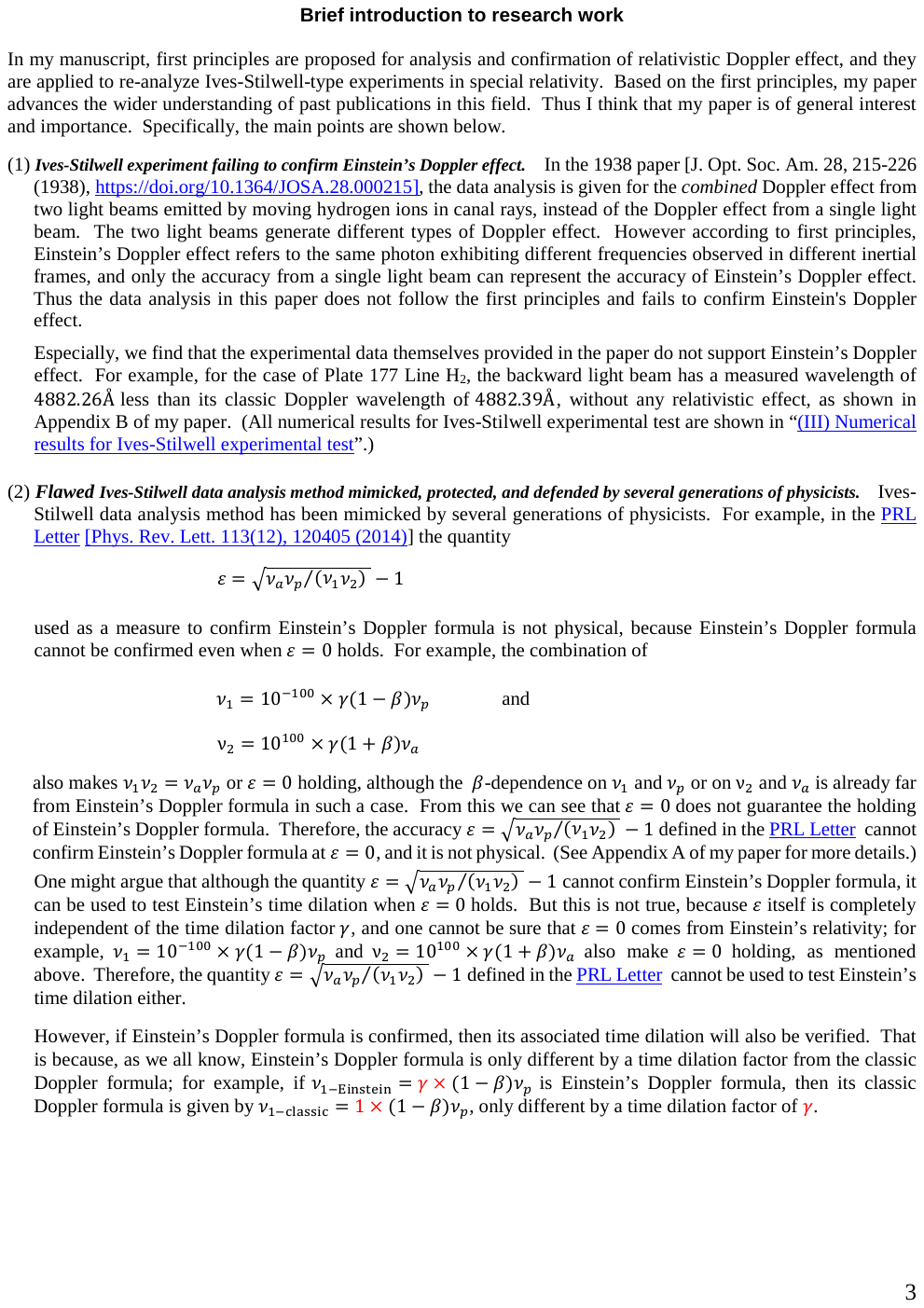}
\label{figM3}
\end{figure} 

\begin{figure} 
\includegraphics[trim=1.0in 1.0in 1.0in 1.0in, clip=true,scale=1.0]{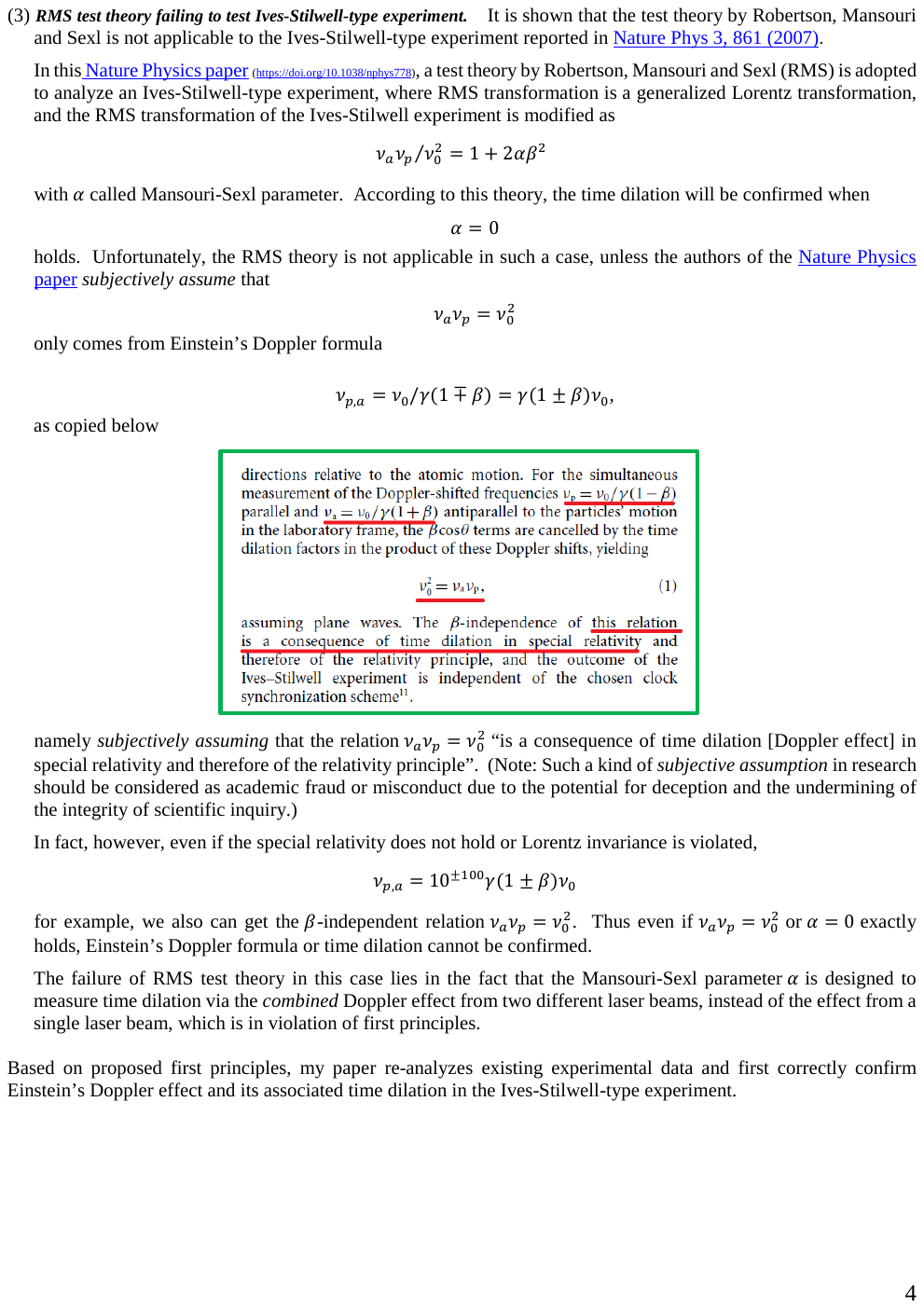}
\label{figM4}
\end{figure} 

\begin{figure} 
\includegraphics[trim=1.0in 1.0in 1.0in 1.0in, clip=true,scale=1.0]{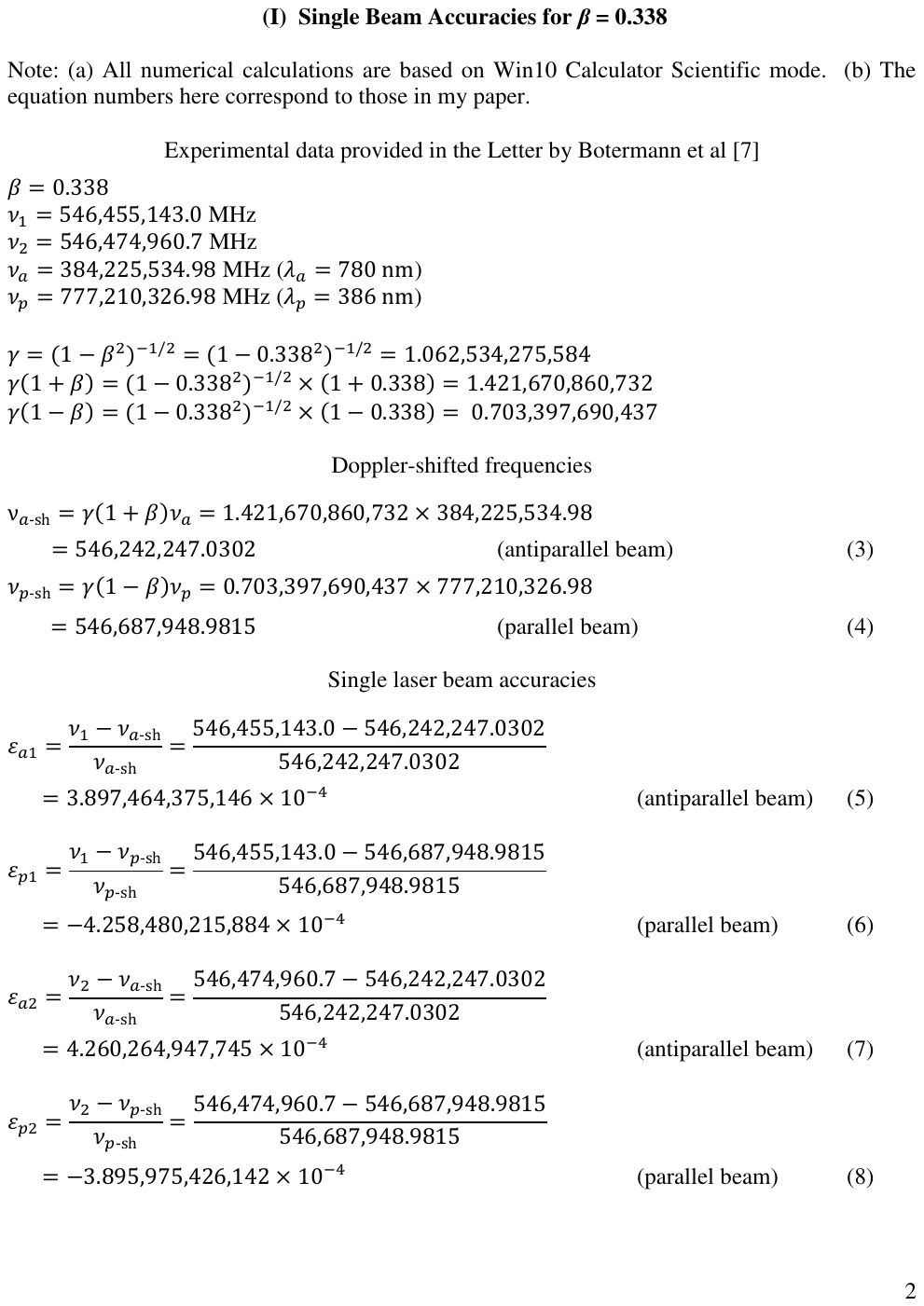}
\label{figM5}
\end{figure} 

\begin{figure} 
\includegraphics[trim=1.0in 1.0in 1.0in 1.0in, clip=true,scale=1.0]{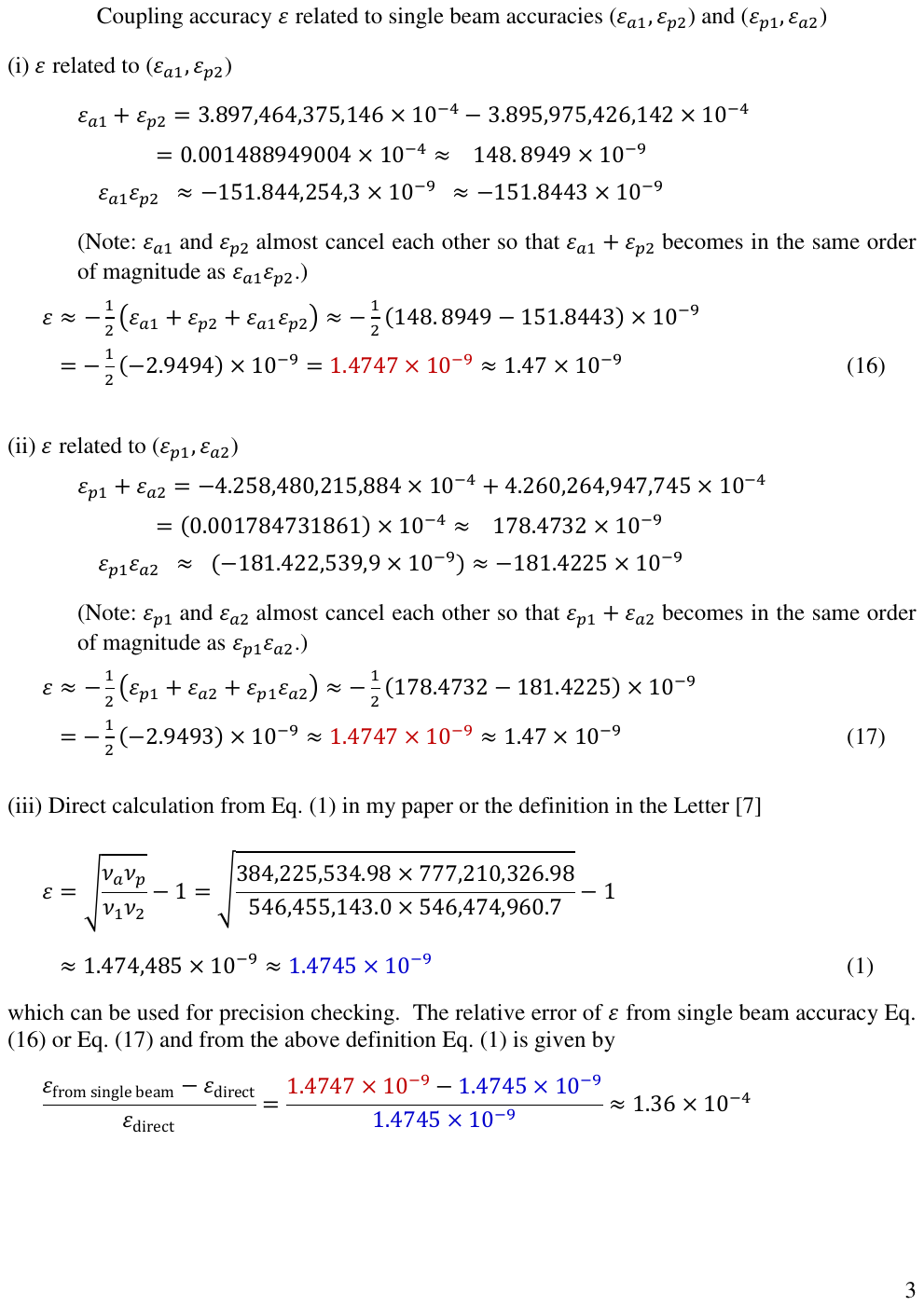}
\label{figM6}
\end{figure} 

\begin{figure} 
\includegraphics[trim=1.0in 1.0in 1.0in 1.0in, clip=true,scale=1.0]{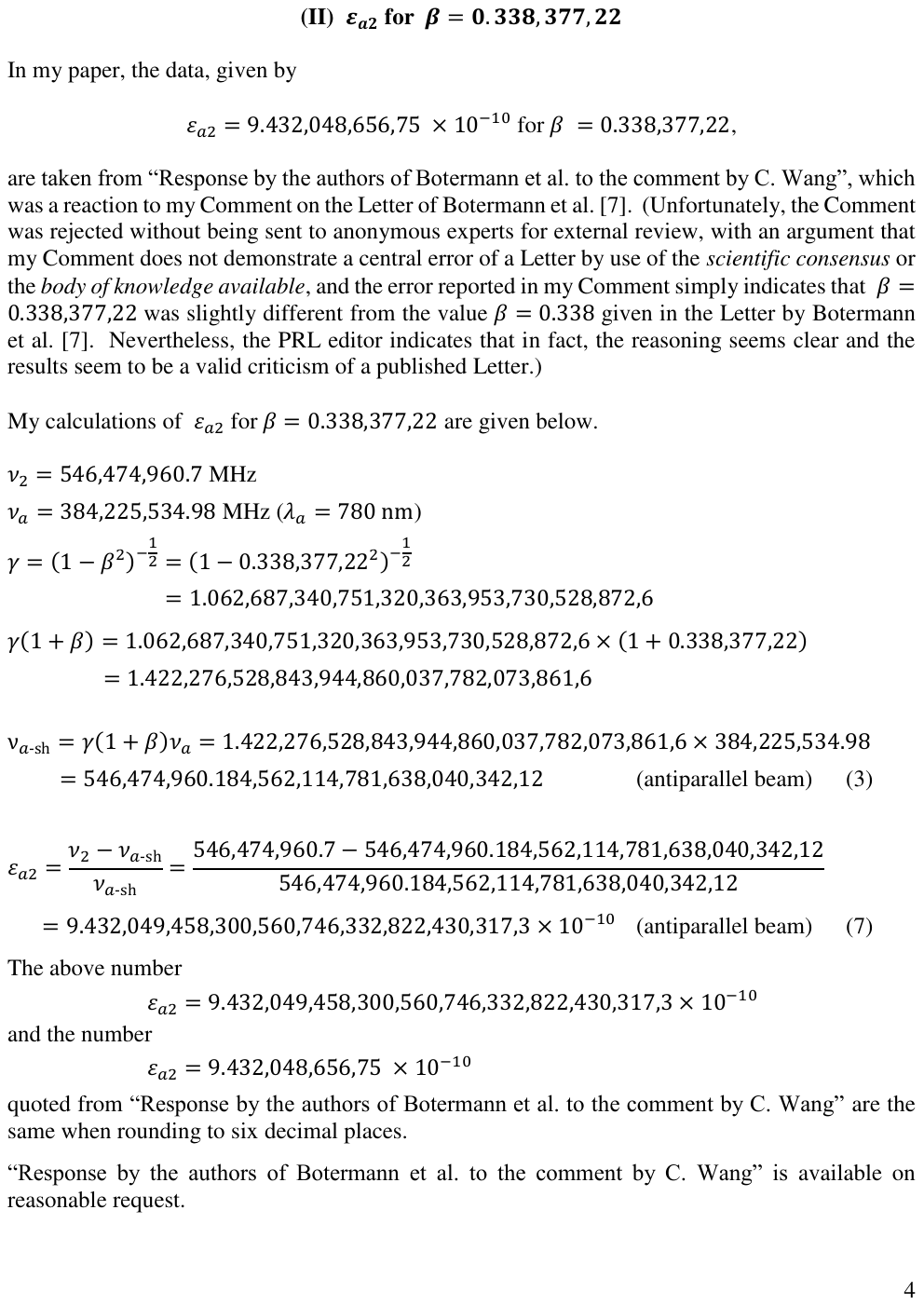}
\label{figM7}
\end{figure} 

\begin{figure} 
\includegraphics[trim=1.0in 1.0in 1.0in 1.0in, clip=true,scale=1.0]{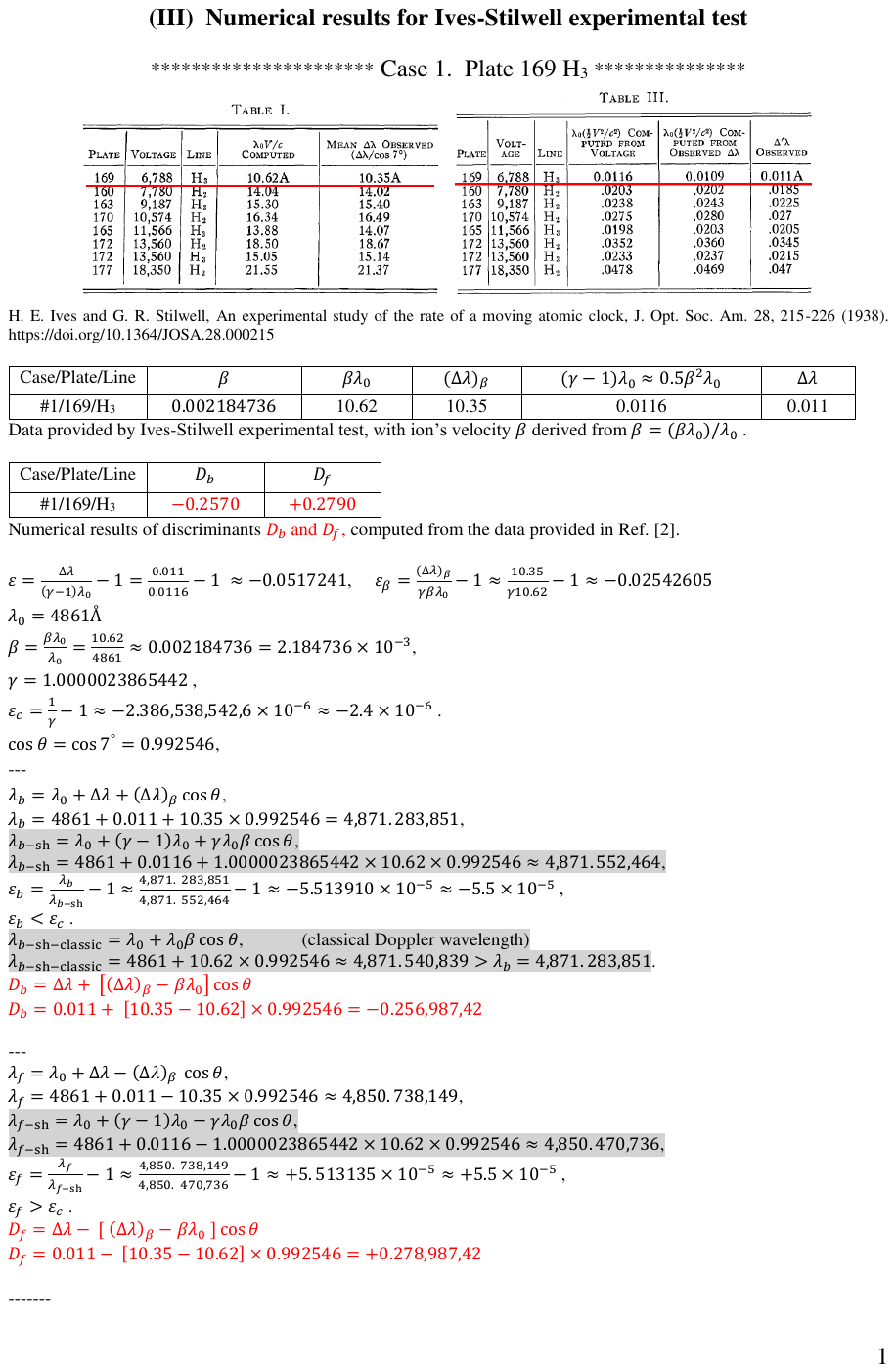}
\label{figM8}
\end{figure} 

\begin{figure} 
\includegraphics[trim=1.0in 1.0in 1.0in 1.0in, clip=true,scale=1.0]{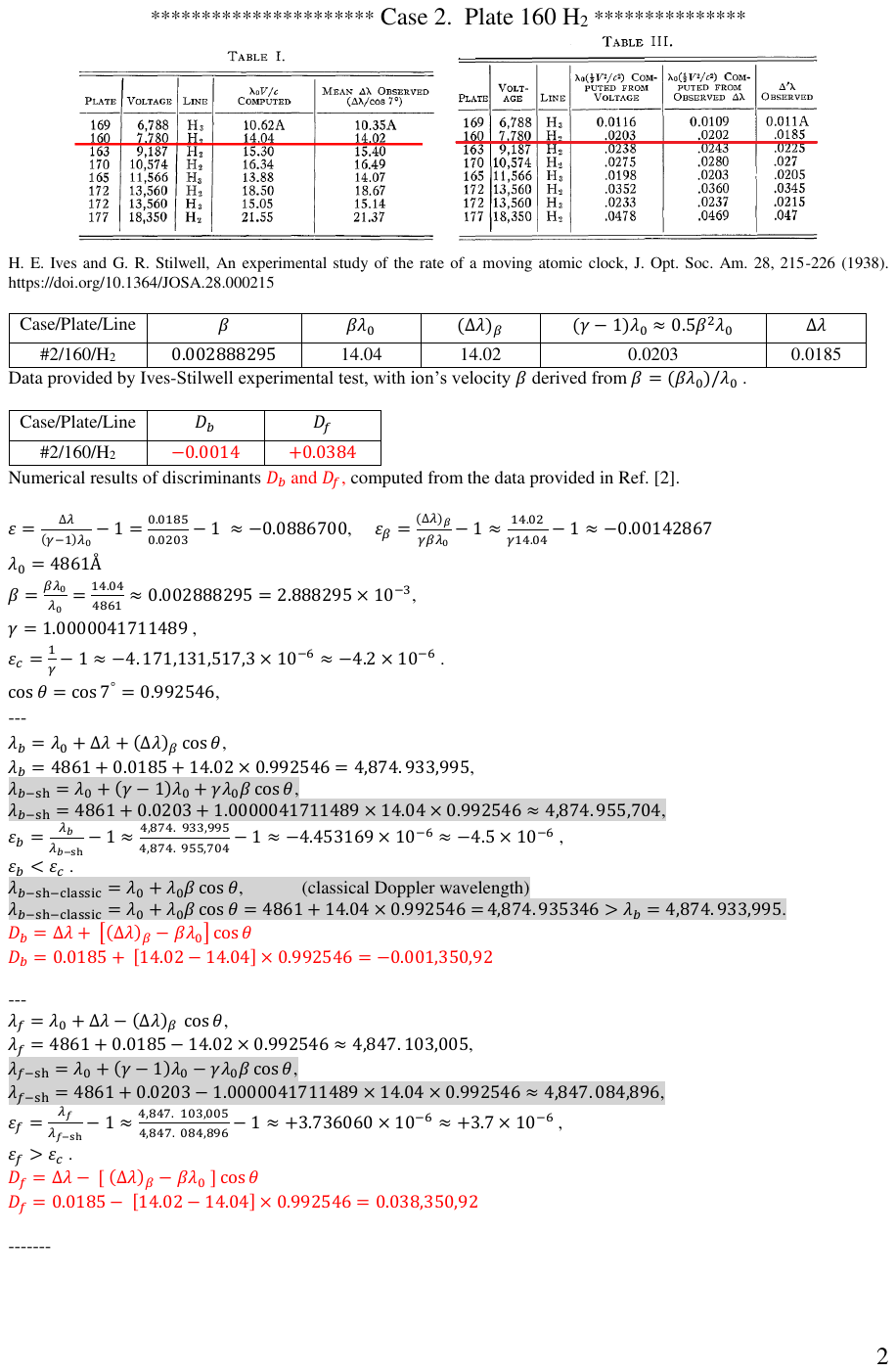}
\label{figM9}
\end{figure} 

\begin{figure} 
\includegraphics[trim=1.0in 1.0in 1.0in 1.0in, clip=true,scale=1.0]{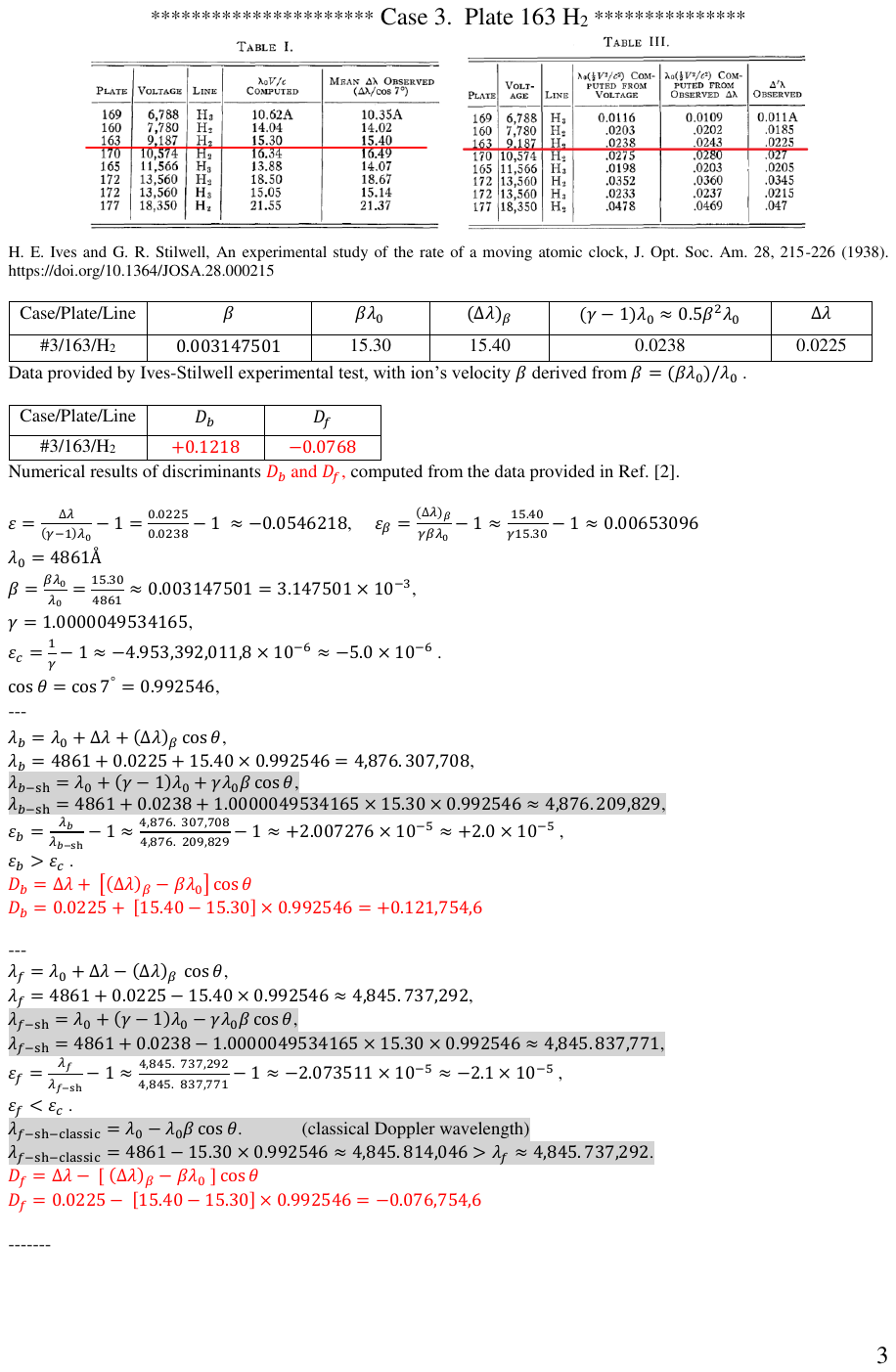}
\label{figM10}
\end{figure} 

\begin{figure} 
\includegraphics[trim=1.0in 1.0in 1.0in 1.0in, clip=true,scale=1.0]{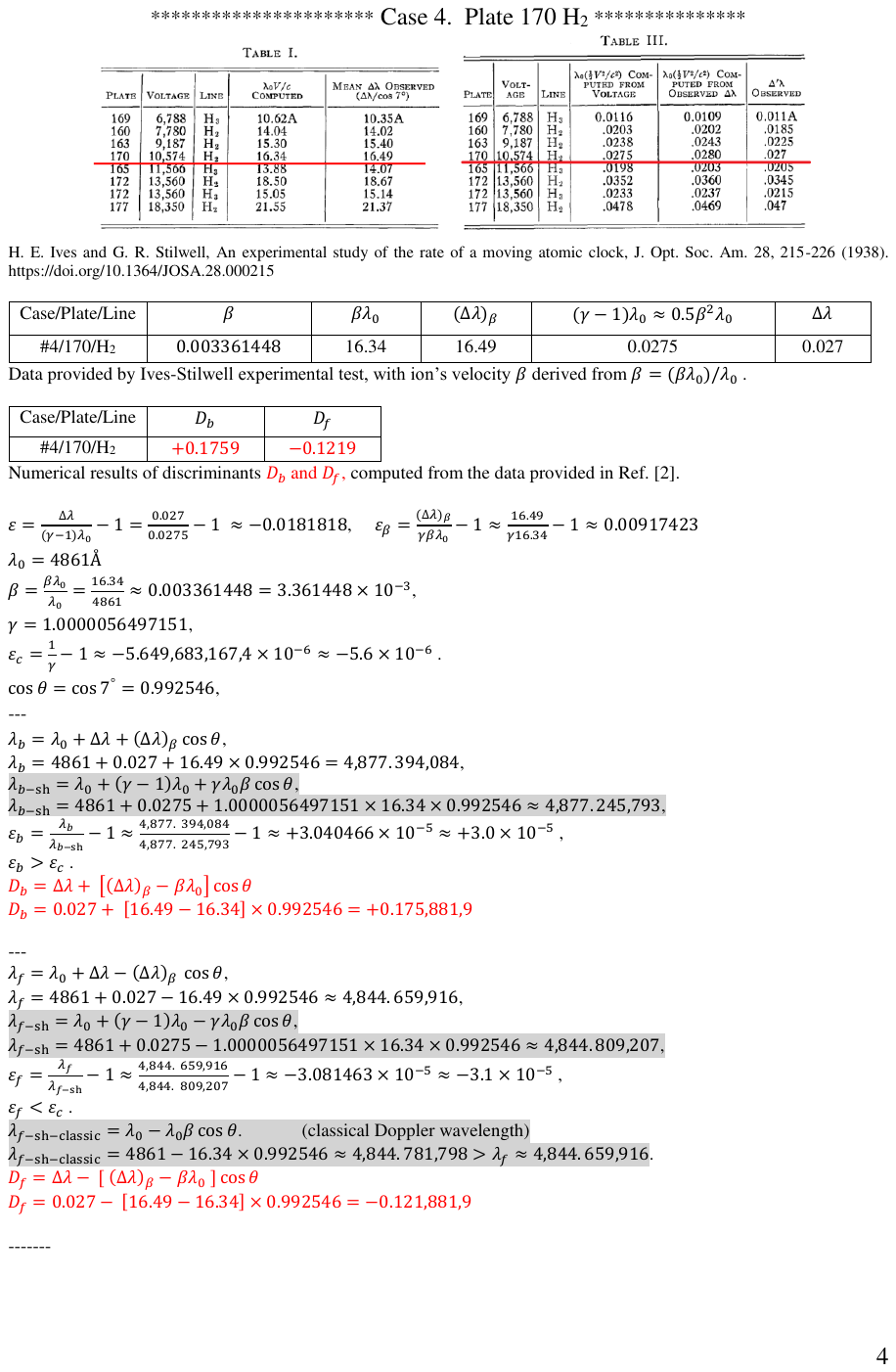}
\label{figM11}
\end{figure} 

\begin{figure} 
\includegraphics[trim=1.0in 1.0in 1.0in 1.0in, clip=true,scale=1.0]{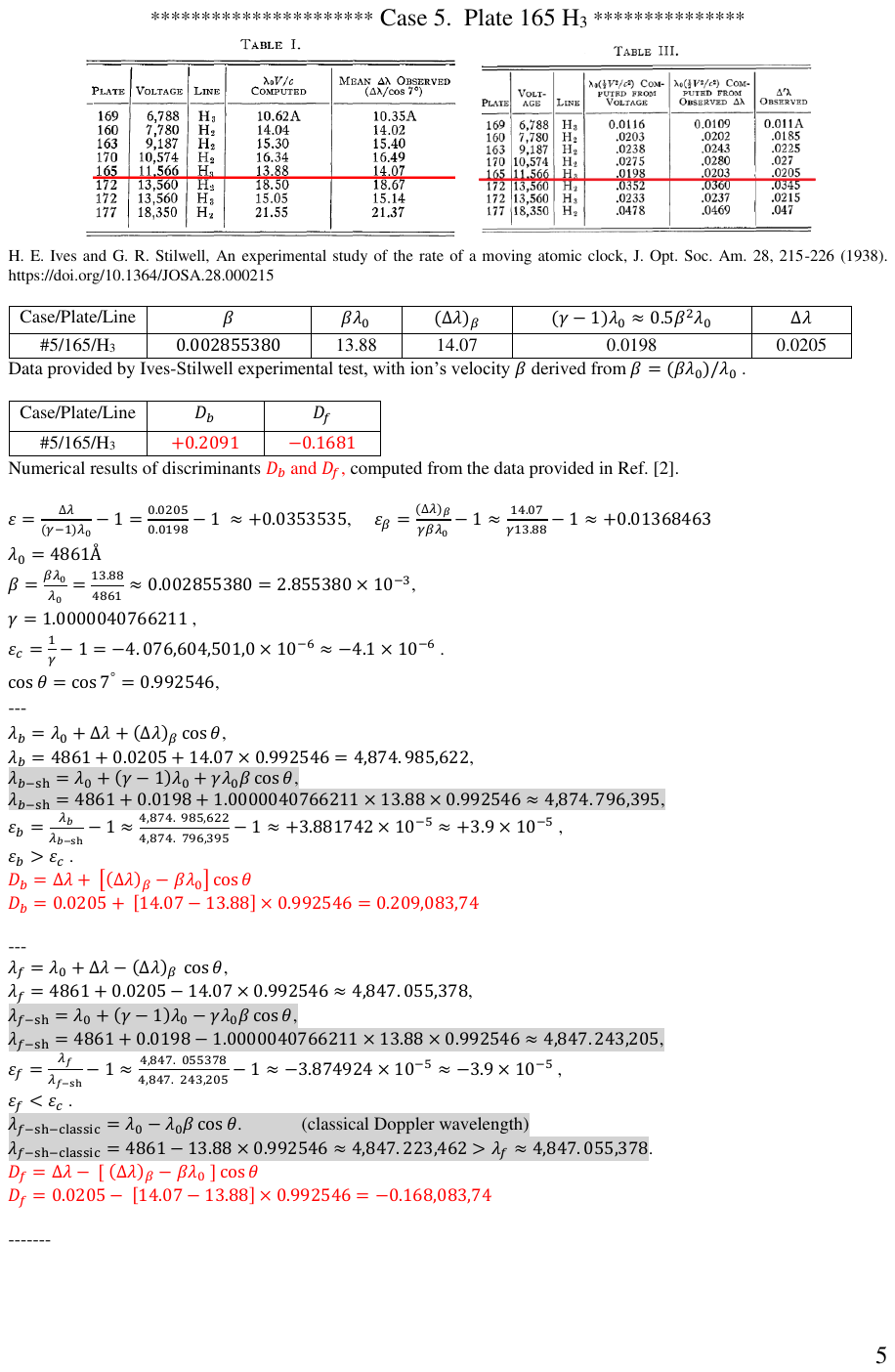}
\label{figM12}
\end{figure} 

\begin{figure} 
\includegraphics[trim=1.0in 1.0in 1.0in 1.0in, clip=true,scale=1.0]{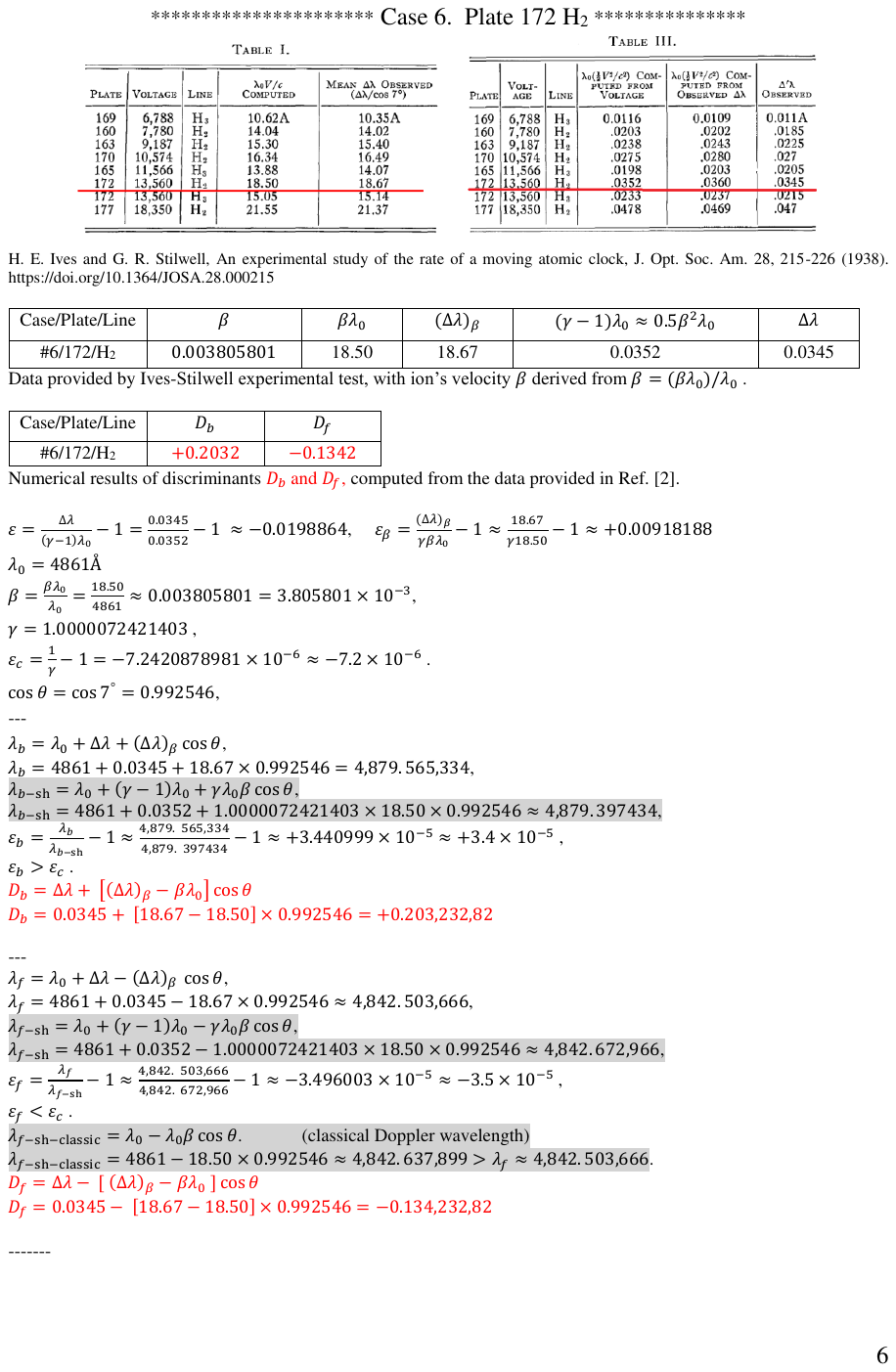}
\label{figM13}
\end{figure} 

\begin{figure} 
\includegraphics[trim=1.0in 1.0in 1.0in 1.0in, clip=true,scale=1.0]{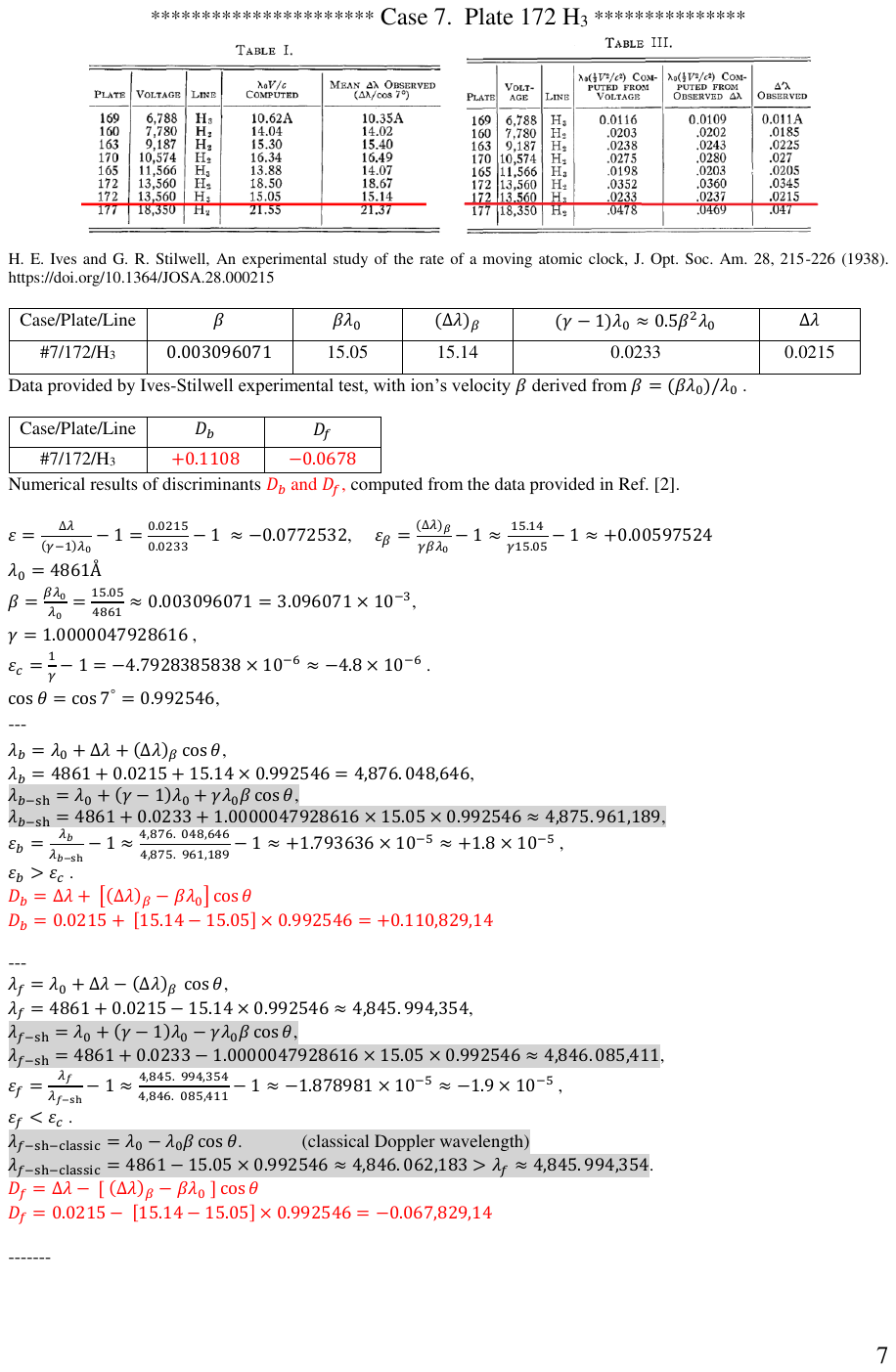}
\label{figM14}
\end{figure}

\begin{figure} 
\includegraphics[trim=1.0in 1.0in 1.0in 1.0in, clip=true,scale=1.0]{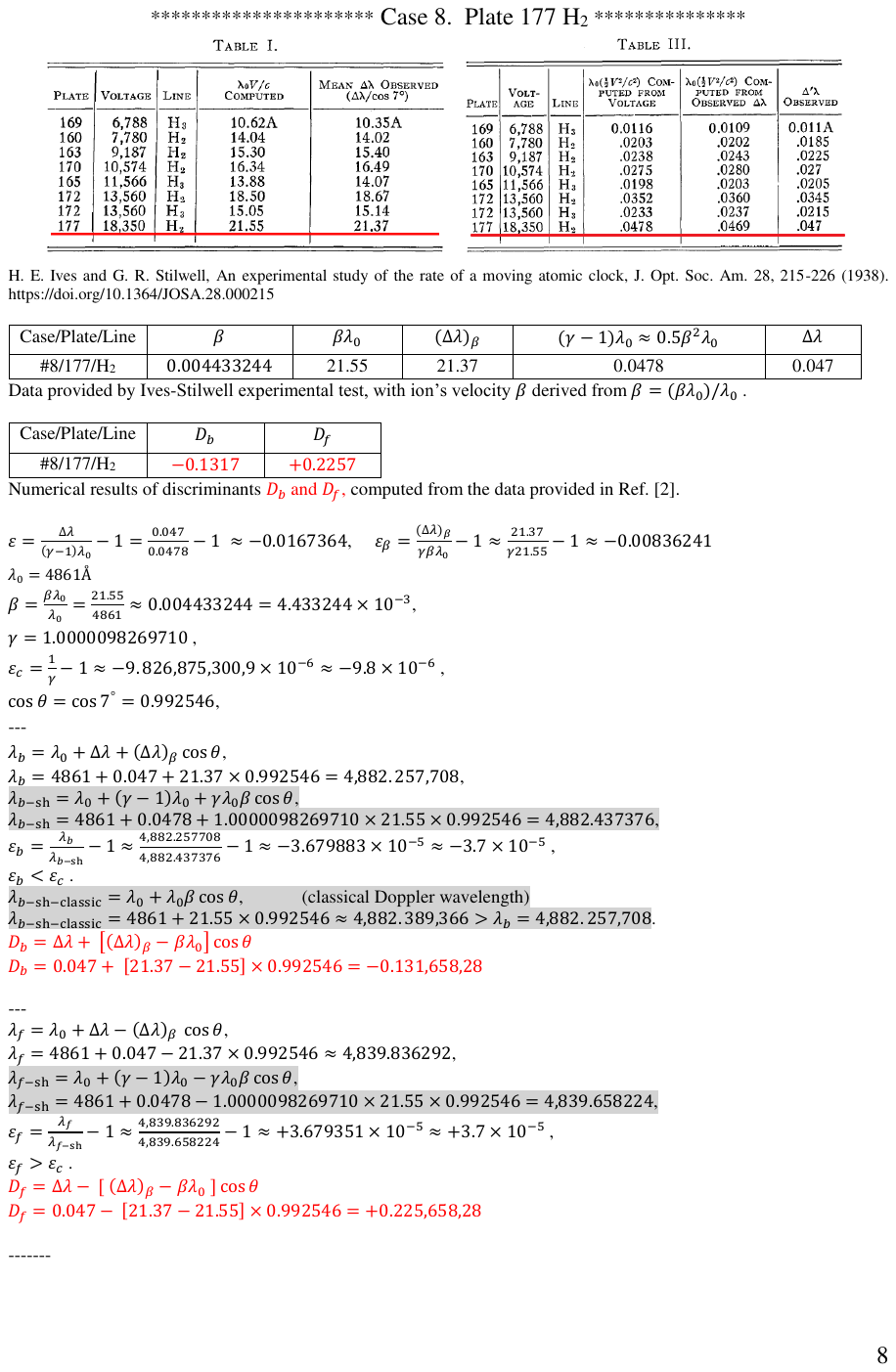}
\label{figM15}
\end{figure} 

\end{document}